\documentclass{aa}
\usepackage{tikz}
\usepackage{graphicx}
\bibpunct{(}{)}{;}{a}{}{,} 
\usepackage[varg]{txfonts}
\usepackage[bookmarks=true,
  colorlinks,
  linkcolor=blue,
  urlcolor=blue,
  citecolor=blue,
  plainpages=false,
  pdfpagelabels,
  final,
  breaklinks=true]{hyperref}
\usepackage{mathtools}

\begin{document}

\title{Magnetic field strength constraints on $\boldsymbol{\gamma}$-ray flaring regions in the flat spectrum radio quasar PKS~1222+216}
 
   \author{Yeji Jo
          \inst{1}
          \and
          Sang-Sung Lee\inst{2,3,\thanks{Corresponding author}}
          }

   \institute{Department of Astronomy and Space Science, Kyung Hee University, 1732, Deogyeong-daero, Giheung-gu, Yongin-si, Gyeonggi-do 17104, Republic of Korea\\
              \email{yejijo@khu.ac.kr}
         \and
             Korea Astronomy and Space Science Institute, 776 Daedeok-daero, Yuseong-gu, Daejeon 34055, Republic of Korea\\
             \email{sslee@kasi.re.kr}
        \and
            Korea University of Science and Technology, 217 Gajeong-ro, Yuseong-gu, Daejeon 34113, Korea\\
             }

   \date{Received June 15, 2025; accepted July 16, 2025}

\authorrunning{Yeji Jo et al.}
\titlerunning{Magnetic field strength constraints on $\gamma$-ray flaring regions in the flat spectrum radio quasar PKS~1222+216}

\abstract
{We present a multi-wavelength study of the Flat Spectrum Radio Quasar PKS~1222+216, analyzing its long-term variability of radio data obtained in 2013--2020 from the iMOGABA, MOJAVE, and VLBA-BU-BLAZAR programs, along with $\gamma$-ray data from \textit{Fermi}-LAT. We found that the radio flux densities at 15, 22, 43, and 86~GHz declined exponentially by 37\%--56\% over a year following a $\gamma$-ray flare in November 2014. We estimated jet physical parameters through Gaussian model fitting of VLBA 43~GHz data, identifying 10 jet components. The cooling timescales of the jet emission regions, i.e., newly ejected components C9, C10, and C11, range from 43 to 222 days, with the estimated jet viewing angles of approximately 8 degrees and magnetic field strengths of 77--134~mG in the jet emission regions. Additionally, by determining the magnetic field strength at different frequencies, we found that the magnetic field scales as $B\propto r^{-0.3\pm0.04}$, indicating a non-equipartition condition ($k_\text{r}\gtrsim 1$)  or a slow decline in magnetic field strength profile ($m<1$). By analyzing component ejection times, we discovered that the $\gamma$-ray flare in 2014 coincided with the interaction between the moving component C9 and the stationary feature A1.
We estimated that the $\gamma$-ray emission region is located at $9.2\pm1.0$~pc from the central engine, beyond the BLR and dusty torus, suggesting that the seed photons for inverse Compton scattering originate from the jet itself, external CMB radiation, or a surrounding sheath. Our results favor a scenario where $\gamma$-ray emission originates further downstream from the central engine through interactions between moving and stationary components.
Additionally, our study presents an alternative method for estimating magnetic field strengths in AGNs undergoing long-term synchrotron cooling based on the associated timescale.}


\maketitle
\section{Introduction}

Active galactic nuclei (AGNs) are central compact regions of the galaxies that emit electromagnetic radiation ranging from radio to $\gamma$-rays. 
AGNs are among the most powerful sources in the universe, containing supermassive black holes (SMBHs) with masses ranging from $\sim 10^6$ to $10^{10} M_\odot$ at their centers, surrounded by an accretion disk, and relativistic jets perpendicular to the disk.
Magnetic fields are believed to be responsible for the collimation and acceleration of relativistic jets throughout their propagation.
These jets emit non-thermal synchrotron radiation, resulting from the interaction of relativistic particles with magnetic fields.
The emission radiated from the jets is highly polarized and variable with characteristic timescales of hours to years~\citep{krolik1999active}.
Very long baseline interferometry (VLBI) has been extensively used to study the innermost regions in AGNs, providing high-resolution imaging capabilities.
In particular, utilizing data from the VLBA-BU-BLAZAR monitoring program~\citep{galaxies4040047}, several studies have investigated the temporal evolution of jets or the relationship between $\gamma$-ray variability and jet kinematics~\citep[e.g.,][]{2010AAS...21522502J,3c84,2015A&A...578A.123R}.
The interferometric monitoring of gamma-ray bright AGNs (iMOGABA) employs the Korean VLBI network (KVN) at 22, 43, 86, and 129~$\mathrm{GHz}$~\citep{Lee_2016}, contributing to our understanding of various physical properties of AGNs with bright $\gamma$-rays~\citep[e.g.,][]{lee2020interferometric,kim2022magnetic,park2019ejection}.

Blazars, a subclass of AGN, belong to Type 0 \citep{urry1995unified}, which have very unusual spectral characteristics (including high luminosity and strong variability) with a small viewing angle (e.g., less than a few degrees) of jets relative to the line of the sight, causing substantial Doppler boosting.
A typical spectral energy distribution (SED) exhibits two humps: one attributed to synchrotron radiation at relatively low frequencies, and the other to inverse Compton (IC) radiation at higher frequencies, arising from the scattering of photons, by relativistic electrons, from synchrotron emission or external components (e.g., accretion disk, broad-line region (BLR), or dusty torus).
Blazars are divided into two subclasses using spectral features: flat-spectrum radio quasars (FSRQs) and BL Lac objects (BL Lacs).
FSRQs are highly luminous and radiate strong, with broad emission lines, indicating the presence of a dense BLR, while BL Lacs show weak or absent emission lines.
Magnetic fields in jets play a crucial role in shaping the physical properties of blazars, influencing both jet formation and high-energy emission processes~\citep{marscher2008inner}.
\citet{sokolovsky2010constraining} showed that the upper limits of the magnetic field strength in blazars range from $10^{-2}$ to $10^{2}~\mathrm{Gauss}$.
Many studies have measured the magnetic field strengths of blazars using various methods, including those based on the synchrotron self-absorption (SSA) spectrum~\citep{lee2017interferometric, kang2021interferometric, jeong2023double, kim2022magnetic}, the core shift effect~\citep{o2009magnetic, kutkin2014core}, and the synchrotron luminosity \citep{kim2017millimeter}.
Blazars exhibit $\gamma$-ray flares when superluminal jet components pass through the core, or stationary features, which are often associated with recollimation shocks within the jet \citep{marscher2008inner,kovalev2009relation, li2024multi}. 
The studies have shown that these $\gamma$-ray flares occur at locations far from the central engine (e.g., $\gtrsim 1$~parsec), while the origin of the seed photons responsible for the emission remains under debate.

PKS~1222+216 (also known as 4C~+21.35) is a FSRQ with redshift $z=0.435$ \citep{wang2004connection}, located at $\alpha=12^h24^m54.458^s,~\delta=+21^\circ22’46.388’$ in the $\mathrm{J}2000$ position. 
Its luminosity distance is $D_L= 2.4~\text{Gpc}$ {\citep{jorstad2014jet}}, and the central SMBH of the source has a mass of $6\times10^8~\mathrm{M_\odot}$ \citep{farina2012optical}.
The source has been monitored by Fermi-LAT, and $\gamma$-ray flaring events were detected in 2010, 2014, and 2019~\citep{kramarenko2022decade}.
Among these events, the 2010 flare was identified at very high energies (VHE; $E> 100~\mathrm{GeV}$) by MAGIC \citep{aleksic2011magic}.
\cite{jorstad2014jet} asserted that the $\gamma$-ray maxima coincide with interactions between moving jet knots and stationary structures, such as the core and A1, which are interpreted as recollimation shocks.
They considered that the absence of a spectral cutoff constrains the $\gamma$-ray emission region, causing it to lie outside the BLR to avoid the pair-production absorption of VHE $\gamma$-rays.
This supports the study presented in \cite{tavecchio}, where the emission from a compact blob and a large region located beyond the BLR adequately explains the SED during the 2010 $\gamma$-ray flaring period. 
\cite{lee2019jet} analyzed the trajectory of a newly identified component, C, at 22~GHz using the KVN and VLBI Exploration of Radio Astrometry (VERA) array (KaVA) from 2014 to 2016, and suggested that component C may be associated with the $\gamma$-ray flare observed in 2014.
Kinematic analyses of the jet components in PKS~1222+216 have been extensively conducted using VLBI imaging techniques.
\cite{galaxies4040072} identified five superluminal knots ($9$–$22~c$) and a stationary component A1 ($0.14\pm0.04$~milliarcseconds~(mas) from the core) using 43 GHz VLBA data from 2008 to 2015, including the emergence of knot K6 in 2015.
\cite{lee2019jet} analyzed a subsequent epoch (2014-2016) using KaVA, presenting the kinematics of two jet knots with apparent speeds of $3.5~c$ to $6.8~c$ at 43~GHz.
\cite{Weaver_2022} analyzed the motion of jet components at 43~GHz for PKS~1222+216 using data collected from 2007 to 2018.
The study showed physical parameters for each component, including the timescale of variability, Doppler factor ($\delta$), Lorentz factor ($\Gamma$), and viewing angle. 
Fourteen components were identified, with the parameters for components B9, B10, and B11 estimated as $\delta \sim 3$--6, $\Gamma \sim 3$--13, and with a viewing angle of approximately $6^\circ$.
In addition, several studies have reported estimates of the magnetic field strength.
\cite{pushkarev2012mojave} applied core-shift measurements between 8~GHz and 15~GHz to probe the physical conditions in AGN jets.
For PKS~1222+216, they estimated the magnetic field strength to be 0.9~G at a distance of 1~pc from the central engine, and inferred the location of the 15~GHz core to be 23.41~pc from the central engine.
\cite{lee2019jet} found that the decay timescales of the jet flux densities at 22 and 43~GHz ranged from $\sim$~0.5~yr to 3~yr, resulting in the magnetic field strength $B$ of about 10--30~mG for PKS~1222+216.

In this paper, we present observational results based on long-term, multi-frequency (22–86 GHz) KVN data obtained from 2013 to 2020, supplemented by VLBA radio data, with a focus on constraining the magnetic field strength under synchrotron cooling conditions and investigating the relationship between $\gamma$-ray flares and radio emissions.
In Section~\ref{2}, we describe observation details and how to reduce the data.
Our analysis results are shown in Section~\ref{3}, and our results are discussed in Section~\ref{4}.
Finally, we summarize our study in Section~\ref{5}.

\begin{figure*}[htb!]
\centering
\includegraphics[width=0.8\linewidth]{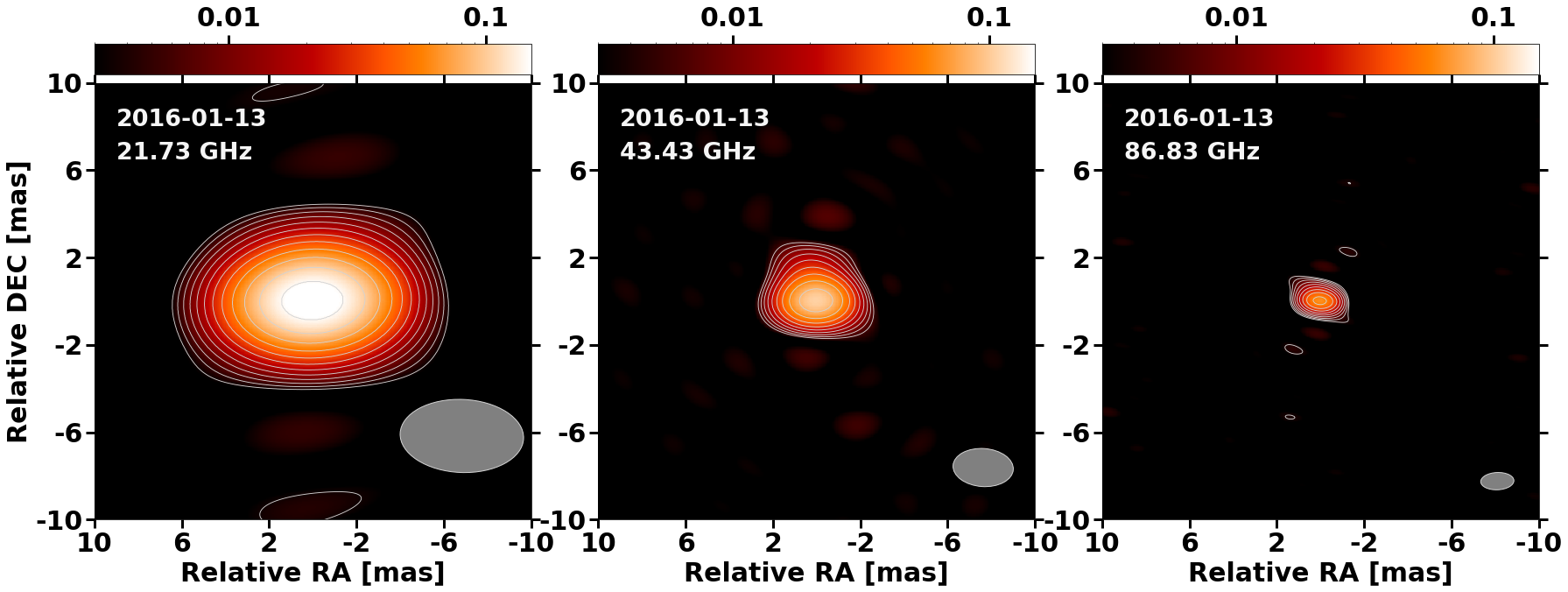}
\caption{ CLEAN images of PKS 1222+216 obtained by KVN observations at 22, 43, and 86~GHz on January 13, 2016. The radio cores are aligned to the (0,0) position. Color and contours indicate the intensity of each map. The contours start at three times the root mean square (RMS) of the residual map and increase by a factor of 1.4. The RMS levels are 0.012, 0.019, and 0.013~Jy/beam at 22, 43, and 86 GHz, respectively. Synthesized beams are plotted on the lower right side of each map as a gray ellipse.}\label{fig:fig1}
\vspace{5mm}
\end{figure*}

\begin{figure*}
\centering
\includegraphics[width=1.0\linewidth]{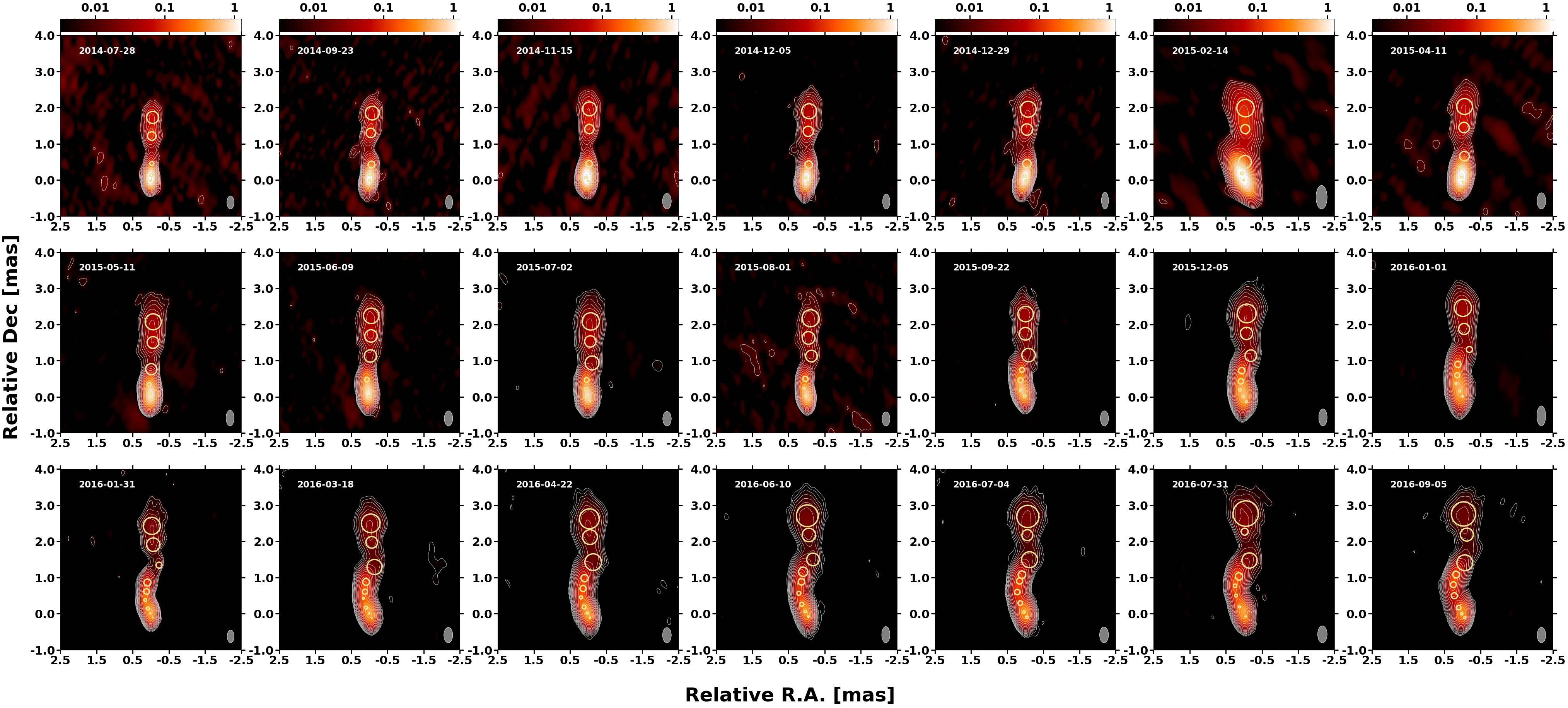}
\caption{CLEAN maps at VLBA 43~GHz from July 2014 to September 2016. The contour starts from three times the RMS noise level and increases by a scale factor of 1.4. The yellow circles indicate the sizes (FWHM) of individual components at each observation. The gray ellipses indicate the synthesized beam sizes for each epoch.} \label{fig:fig5}
\vspace{5mm} 
\end{figure*}

\begin{figure}[htb!]
\centering
\includegraphics[width=\columnwidth]{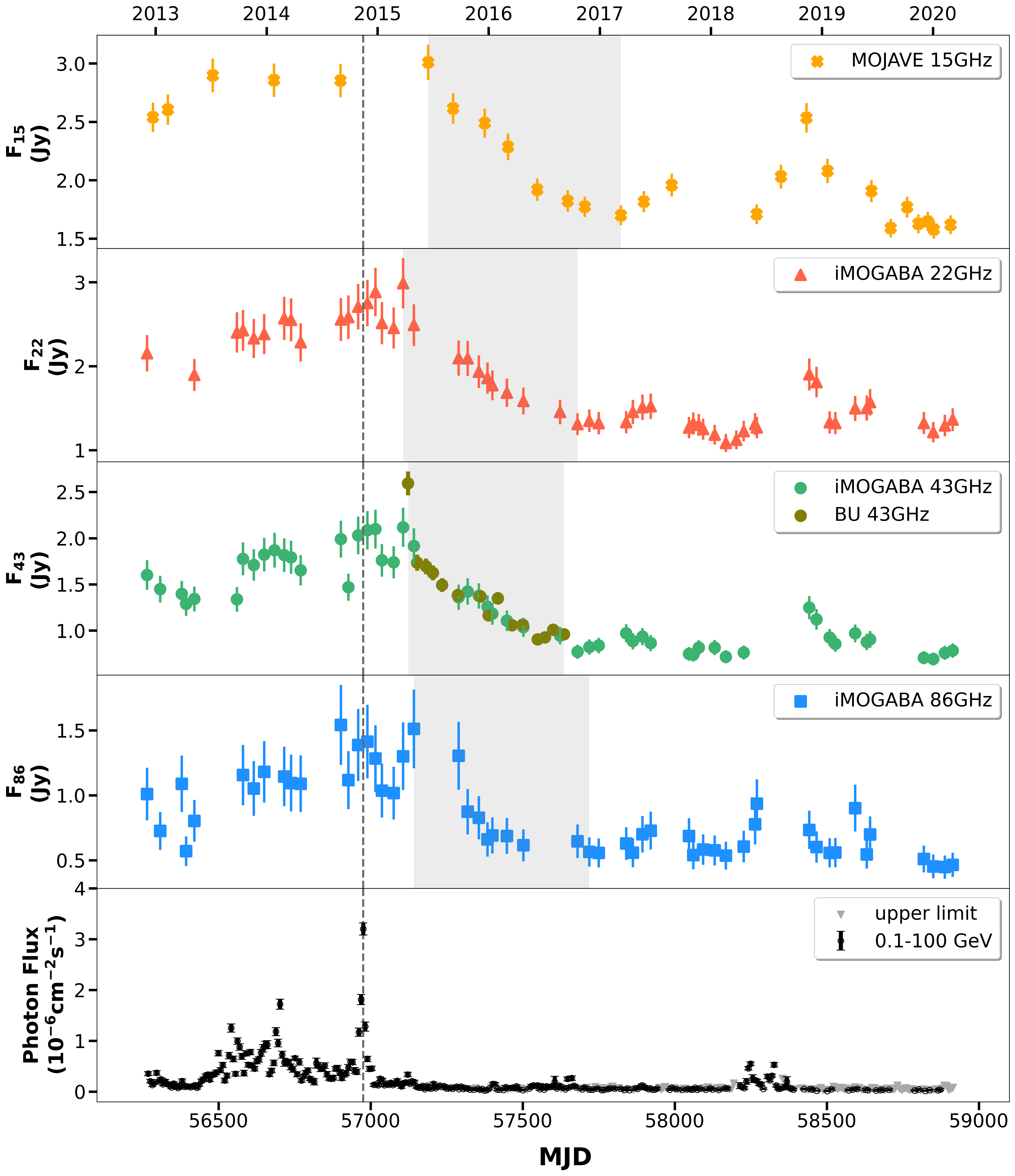}
\caption{
Light curves of PKS 1222+216 observed from December 2012 to March 2020 (MJD~56265--58914) at 15, 22, 43, and 86~GHz. The total CLEAN flux densities are shown for each frequency.
In the third panel, the green circles represent the iMOGABA data, while the olive circles correspond to the BU data. The error bars used typical values of 10\% for K and Q, and 20\% for W. BU and MOJAVE were applied at 5\%~\citep{jorstad2017kinematics}. The bottom panel shows the 0.1--100~GeV $\gamma$-ray light curve from the Fermi-LAT LCR. The upper limit of the flux density is indicated by the gray triangles. The gray vertical dashed line refers to the $\gamma$-ray peak at MJD~56975.5~\citep{ezhikode2022long}. The gray boxes are the periods that fit the exponential decay profiles at each frequency.
} \label{fig:fig2}
\vspace{5mm}
\end{figure}

\section{Observations and multi-wavelength data}\label{2}

\subsection{KVN observations and data reduction}
We obtained multi-frequency simultaneous KVN data for the target source PKS 1222+216 as part of the KVN key science program (iMOGABA). 
The observations were conducted at 22 (K), 43 (Q), 86 (W), and 129 GHz (D) bands utilizing the KVN Yonsei, Ulsan, and Tamna stations.
The observations were conducted monthly, except during maintenance seasons from June to August every year.
The full bandwidth of 256~MHz was divided into four frequencies, providing a bandwidth of 64~MHz for each frequency in left circular polarization. The source was observed for 1--8 scans of 5 minutes in length, distributed over several hours. Further detailed observations can be found in \cite{Lee_2016}.
The data presented here cover the period from December 4, 2012 (MJD~56265) to March 6, 2020 (MJD~58914). 
A total of 60 epochs were obtained at K and Q bands, while 63 and 41 epochs were available for W and D bands, respectively.
The KVN data were processed by the DiFX software correlator at the Korea Astronomy and Space Science Institute (KASI)~\citep{lee2014early}.
The data were also calibrated with the AIPS package utilizing an automatic pipeline for KVN data~\citep{hodgson2016automatic}, including the procedures of phase and amplitude calibrations, fringe fitting, and bandpass correction. 

We used the pipelined KVN data to obtain CLEAN images via the \texttt{DIFMAP} software~\citep{difmap_1994}. 
Amplitude outliers were flagged, and the imaging process began with the self-calibrating of the visibility data based on a point-source model of 1~Jy using the \texttt{DIFMAP} command \texttt{STARTMOD}. 
Subsequently, CLEANing and phase self-calibration were performed iteratively until the residual maps were left with only Gaussian random noise, i.e., a uniform distribution of residual noise with comparable positive and negative maxima.
We obtained CLEAN images of the target source in 54 epochs on the K and Q bands, in 53 epochs on the W band, and 23 epochs on the D band.
The number of images on the D band is less than half that on the K, Q, and W bands due to the lack of baselines; accordingly, we used only the K, Q, and W bands for the data analysis.
The typical amplitude calibration errors applied here are 10\% for K and Q bands and 20\% for the W band.
Fig.~\ref{fig:fig1} shows the CLEAN image from the KVN data reduced for the 2016 January epoch.
The CLEAN beam sizes are typically 5.66~mas$\times$3.35~mas, 2.77~mas$\times$1.74~mas, and 1.52~mas$\times$0.79~mas at 22, 43, and 86~GHz, respectively.

\subsection{VLBA data}\label{2.2}
PKS~1222+216 has been observed at 43~GHz with the VLBA under the VLBA-BU-BLAZAR program \citep{galaxies4040047}, which monitors 34 blazars and three radio galaxies.
We selected 15 VLBA epochs from the period between April 12, 2015 and September 5, 2016 (MJD~57124--57636) to complement the 43~GHz flux density measurements obtained with the KVN.
The VLBA 43~GHz data were pre-calibrated with CLEAN models available on the program website.\footnote{\url{https://www.bu.edu/blazars/VLBAproject.html}}
We obtained the CLEAN total flux densities of PKS 1222+216 from these CLEAN models.
Typical amplitude calibration uncertainty is estimated to be 5\%. 
For kinematic analysis, we also selected 21 VLBA epochs spanning from July 28, 2014 to September 5, 2016 (MJD~56866--MJD~57636) to investigate the inner region of the source, taking advantage of the superior angular resolution of 0.1~mas~\citep{jorstad2017kinematics}.
We fitted circular Gaussian models to the data using the \texttt{MODELFIT} task of \texttt{DIFMAP}, taking into account the findings of previous studies~\citep{jorstad2014jet, galaxies4040072, Weaver_2022}.
We iteratively added Gaussian components to the model, accepting each new component only if it yielded a $\chi^2$ reduction of at least 10\%. 
To maintain continuity, the final model from the previous epoch was used as the initial guess for the subsequent epoch.
Fig.~\ref{fig:fig5} shows the results of the component model fitting, where the yellow circles represent the full width at half maximum (FWHM) size of each component.

The 15~GHz flux density of PKS~1222+216 was obtained from the Monitoring Of Jets in AGN with the VLBA Experiments (MOJAVE) program, referring to the long-term and systematic monitoring of relativistic jets in AGN on parsec scales using VLBA \citep{lister2018mojave}.
We selected eight VLBA observations taken from June 16, 2015 to March 11, 2017 (MJD~57189--57823). 
We utilized the total CLEANed Stokes I flux density in the VLBA image at 15~GHz, publicly available on the program website.\footnote{\url{https://www.cv.nrao.edu/MOJAVE/allsources.html}} 
The typical amplitude calibration error was estimated to be 5\%.

\subsection{$\boldsymbol{\gamma}$-ray data}
The Fermi Large-Area Telescope (LAT) Light Curve Repository (LCR) provides $\gamma$-ray light curves binned on timescales of three days, one week, and one month for 1525 sources, including PKS~1222+216 \citep{abdollahi2023fermi}.\footnote{\url{https://fermi.gsfc.nasa.gov/ssc/data/access/lat/LightCurveRepository/}}. 
From this repository, we retrieved the $\gamma$-ray light curve of PKS~1222+216 with a one-week cadence and energy range of 0.1--100 GeV from December 7, 2012 (MJD~56268) to March 6, 2020 (MJD~58914).

\section{Results}\label{3}
\subsection{Multi-wavelength light curves}\label{3.1}
We obtained CLEAN total flux density light curves at 15, 22, 43, and 86~GHz from MOJAVE and iMOGABA, covering a time span of approximately seven~years from December 2012 to March 2020 (MJD~56265--58914), as shown in Fig.~\ref{fig:fig2}.
The source appears compact within the KVN data (see Fig.~\ref{fig:fig1}) and as such it is most likely that the CLEAN total flux is dominated by the emission from the compact region.
To enhance the light curve cadence at 43~GHz, we incorporated the CLEAN total flux densities of the BU 43~GHz data.
We found that the multi-wavelength CLEAN total flux densities exhibited a declining trend at all frequencies from early 2015 to early 2017 during the observation period.
Following the methodology described in Section \ref{3.4}, we determined the decay periods of the flux densities for each wavelength (gray boxes in Fig.~\ref{fig:fig2}).
Table \ref{tab:decay_param} summarizes the range of total flux densities, the decay periods, and fractional flux density decreases for each frequency. 
In the following sections, we focus on a detailed examination of the physical properties during the cooling period.

\begin{table}[ht]
\centering
\caption{Multi-frequency flux density variability and decay properties}
\label{tab:decay_param}
\begin{tabular}{cccc}
\hline
\hline
Frequency & Flux Range & Decay period & Decrease \\
(GHz) & (Jy) & (MJD) & (\%)  \\
\hline
15 & 1.578--3.010 & 57189--57823 & 56.48  \\
22 & 1.086--2.989 & 57107--57680 & 43.88  \\
43 & 0.693--2.593 & 57124--57636 & 37.03  \\
86 & 0.448--1.542 & 57142--57719 & 51.12  \\
\hline
\end{tabular}
\tablefoot{
The columns represent the observing frequency, the range of total flux density (min--max), the MJD range of the decay period, and the percentage decrease in flux density during this period, respectively.
}
\end{table}

We also utilized $\gamma$-ray data ranging from MJD~56268 to MJD~58914, a period coinciding with the iMOGABA observations (including the upper limits).
The light curve peaked on MJD~56975.5 with a peak flux density of $3.2\times10^{-6}\pm1.20\times10^{-8}$~$\mathrm{photon}~\mathrm{cm}^{-2}\mathrm{s}^{-1}$ and a MJD flaring period of~56974.6--56977.8, as defined in \cite{ezhikode2022long}.
We found that after the $\gamma$-ray flare, the radio light curves peaked at MJD~57189 (15~GHz), MJD~57107 (22~GHz), MJD~57124 (43~GHz), and MJD~57142 (86~GHz), followed by a monotonic decline, as described above.
After a significant flare in 2014, the flux densities began to rise again during 2018--2019, peaking initially at 86~GHz, in agreement with the $\gamma$-ray data. 
This intriguing behavior could be explored further in future studies.

\begin{figure}
\centering
\includegraphics[width=\columnwidth]{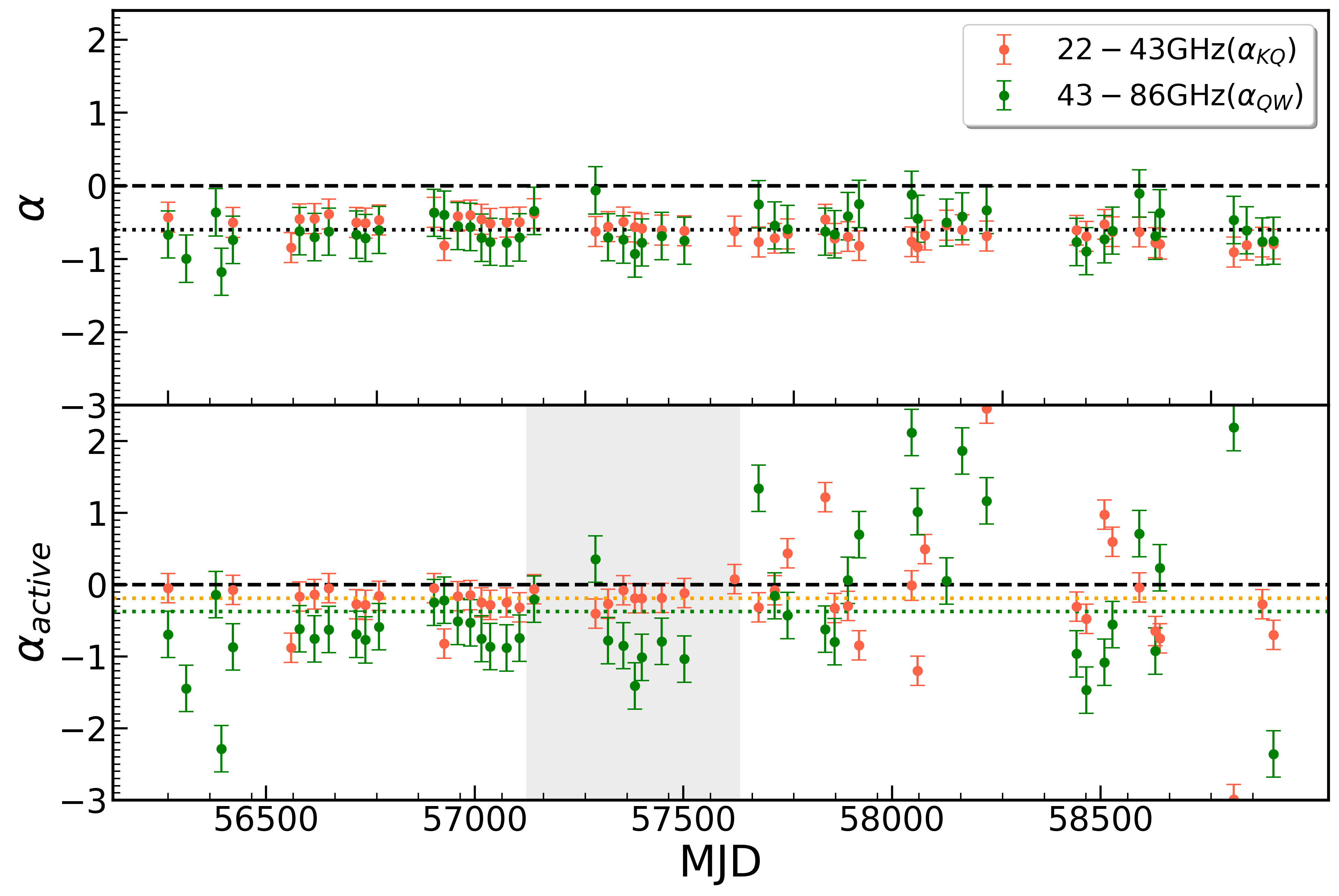}
\caption{Radio spectral indices at 22--43~GHz and 43--86~GHz. The top panel represents the values derived from the CLEAN total flux, while the bottom panel corresponds to the active flux.
The orange scatters represent the spectral index at 22--43~GHz, while the green ones indicate the spectral index at 43--86~GHz.
The horizontal dashed line indicates that the spectral index is zero, and the dotted line of the top panel is the mean value with $\alpha=-0.6$.
In the bottom panel, the orange dotted line represents the average active spectral index of $-0.19$ at 22--43~GHz, while the green dotted line represents the average active spectral index of $-0.37$ at 43--86~GHz.
The gray box indicates the decreasing period from 57124 to 57636 at 43~GHz.} \label{fig:fig3}
\vspace{5mm} 
\end{figure}

\subsection{Spectral indices of the iMOGABA flux density}\label{3.2}

We analyzed multi-wavelength iMOGABA data to investigate the spectral properties of PKS~1222+216 and derived spectral index values from simultaneous KVN observations.
We obtained the spectral index $\alpha$ at two adjacent frequencies (frequency $\nu_1$ and $\nu_2$) as 

\begin{equation}
\label{eq1}
\alpha = \frac{\log(F_{\nu_2}/F_{\nu_1})}{\log(\nu_2 / \nu_1)},
\end{equation}
where $F_\nu$ is the CLEAN flux density at the observing frequency at $\nu$ and $\nu_2, \nu_1$ ($\nu_2 > \nu_1$) are the observing frequencies.

\begin{figure}
\centering
\includegraphics[width=\columnwidth]{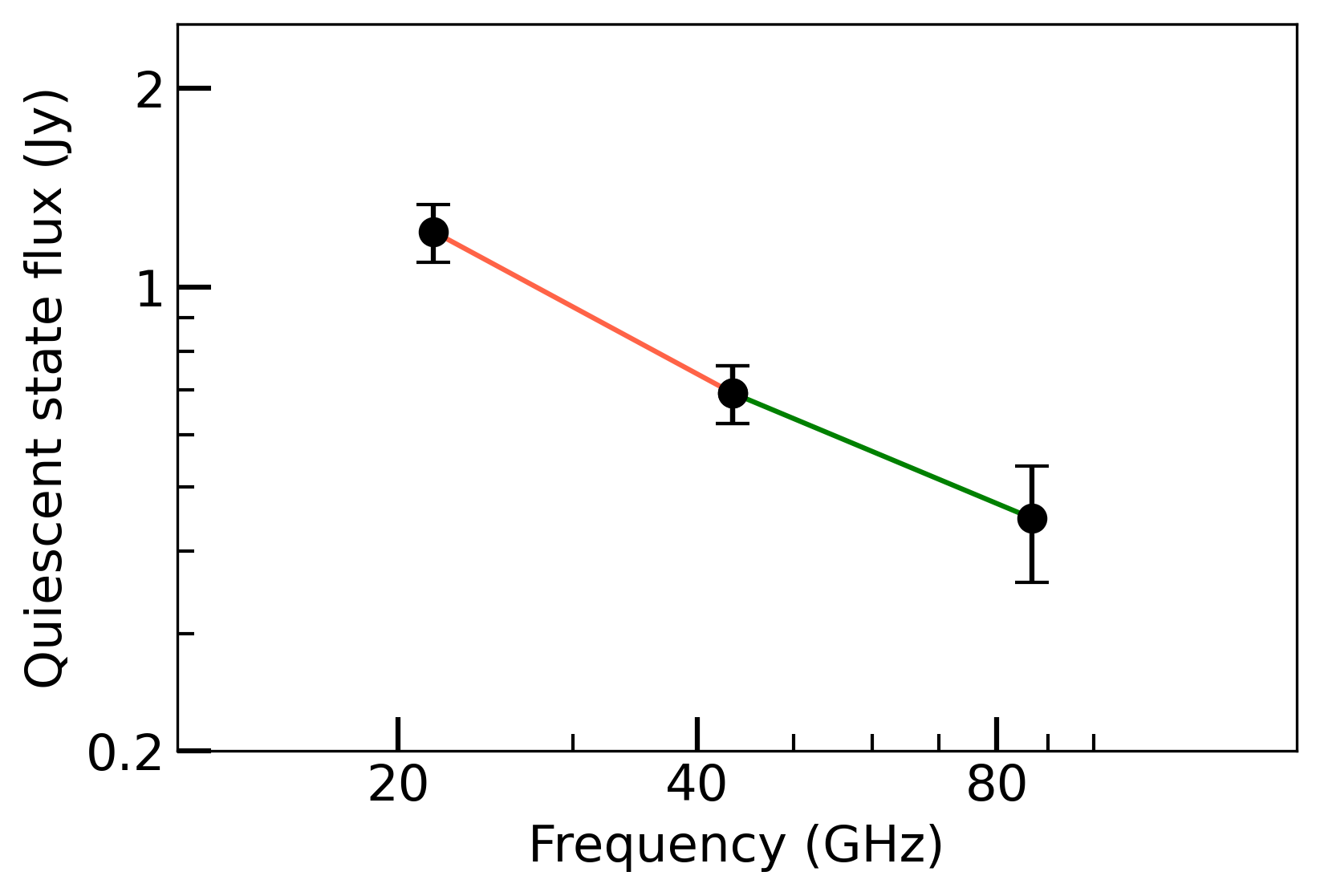}
\caption{Spectral index of PKS~1222+216 in the quiescent state. The black points represent the quiescent-state flux densities at 22, 43, and 86~GHz, with error bars indicating flux uncertainties. The orange slope corresponds to the quiescent spectral index between 22 and 43 GHz ($\alpha_\text{KQ} = -0.81$), while the green slope represents the quiescent spectral index between 43 and 86 GHz ($\alpha_\text{QW} = -0.628$).} \label{fig:fig4}
\vspace{5mm} 
\end{figure}

The top panel of Fig.~\ref{fig:fig3} shows the spectral indices for 53 epochs with corresponding average values of $-0.60\pm0.15$ and $-0.59\pm0.23$ at 22--43~GHz and 43--86~GHz, indicating that the source is optically thin at 22--86~GHz on the mas-scale.
To explore the spectral properties of the variable emission regions, we analyzed the source in both quiescent and active states. 
The quiescent state refers to periods without significant flux enhancements, and the quiescent flux was defined as the minimum CLEAN total flux density observed during the monitoring period: 1.21, 0.69, and 0.45~Jy at the K, Q, and W bands, respectively, on MJD~58850 (K and Q) and 58888 (W).
Fig.~\ref{fig:fig4} shows the quiescent spectral index.
To account for the effect of quiescent states, we subtracted the quiescent flux from the total CLEAN flux to obtain the active flux, from which the active spectral index was subsequently derived.
The active spectral indices were measured and found to be $-0.19\pm0.71$ at 22--43~GHz and $-0.37\pm0.97$ at 43--86~GHz (see the bottom panel of Fig.~\ref{fig:fig3}). 
We found that the active spectral index at 43--86~GHz increased between MJD~57076 and MJD~57289, transitioning from optically thin to optically thick conditions, before returning to optically thin conditions ($\alpha = -0.78$) on MJD~57319.
This dynamic behavior may be due to the kinematic interaction of the 43~GHz components, but a detailed investigation should be conducted in future studies.

\subsection{BU 43~GHz component analysis} \label{3.3}

In order to investigate the inner jet region (less than 1~mas) and its physical parameters, we conducted a detailed component analysis using the BU program data.
The components were identified based on the results in \cite{Weaver_2022}, following the assumption that the geometric distance from the core and flux density of each component does not vary rapidly over time (see Appendix~\ref{App2} for a detailed comparison and identification process).
Table~\ref{tab:table1} presents the model fitting parameters for each component.
In Fig.~\ref{fig:fig6}, the results of the component analysis are shown, including the distances from the core and the flux densities for each component.

We identified ten components, including new knots, C9, C10, C11, and C12, which appeared on November 15, 2014 (MJD~56976), April 12, 2015 (MJD~57124), September 22, 2015 (MJD~57287), and July 31, 2016 (MJD~57600), respectively.
We found that the $\gamma$-ray flare peak (see Section \ref{3.1}) nearly coincided with the appearance of C9 with a one-day offset.
C9 appears to correspond to a new component, K6, that was ejected in 2015, as mentioned in \cite{galaxies4040072}, as K6 appeared after the high $\gamma$-ray peak.
Interestingly, the position angle of the inner region (within 1~mas) significantly changed from $-3.95^\circ$ (Epoch 2, A1) to $26.23^\circ$ (Epoch 3, C9) when component C9 was ejected. As shown in Figs.~\ref{fig:fig5} and \ref{fig:fig6}, the jet in the inner region appears to bend eastward after the third epoch, which is associated with the $\gamma$-ray flare.
This agrees with the 43~GHz KaVA results from 2016 by \cite{lee2019jet}, which also revealed a $\sim$$20^\circ$ westward tilt in the inner region (see Figure~3).
We also discovered stationary components A1 and A2 at $0.15\pm0.01$~mas and $0.35\pm0.05$~mas from the core, respectively, and which are denoted by the yellow and light red dotted horizontal lines in Fig.~\ref{fig:fig6}.
These findings are consistent with previous studies \citep{galaxies4040072,Weaver_2022,jorstad2014jet}.
Additionally, the average geometric distances from the core for the components C5, C6, C8, C9, C10, and C11 are $2.32\pm0.35$, $1.78\pm0.36$, $1.05\pm0.44$, $0.65\pm0.39$, $0.63\pm0.36$ and $0.52\pm0.28$~mas, respectively.
The fluxes of C9, C10, and C11 decreased rapidly during the observation period, with corresponding reductions of 4.67\%, 11.11\%, and 11.29\%, during the decay period described in Section \ref{3.4}

\subsection{Decay timescales} \label{3.4}
We found that all light curves at each observation frequency decreased during the observation period.
Following \cite{terasranta1994brightness} and \cite{burbidge1974physics}, we undertook exponential profile fitting to the light curves to determine the logarithmic variability timescale. 
We used the following exponential decay model,
\begin{equation}
\label{eq2}
    F_\nu = A_\nu \exp(-B_\nu t) + C_\nu ,
\end{equation}
where $F_\nu$ is flux density; $A_\nu$, $B_\nu$, and $C_\nu$ are constants; and $B_\nu = 1/\tau_\nu$. 
$\tau_\nu$ is the decay timescale at frequency $\nu$.
The non-linear least square method was utilized in \texttt{curve\_fit} of SciPy, which is a fundamental algorithm for scientific computing in Python~\citep{vugrin2007confidence}.

It is crucial to select the precise observation period during the exponential fitting to obtain accurate physical quantities and ensure reliable fitting results.
We determined the decay period at each frequency using the method described below.
We selected data only from the period of the overall large decrease in the flux densities found during the iMOGABA observation period.
Starting from the peak of the light curves, the data were grouped, including four epochs subsequently, after which the group's standard deviation and mean value were calculated.
Data grouping was stopped when the mean value of the current group exceeded the mean value of the previous group. 
The decay period ended at the epoch with the minimum value in the final group.

\begin{figure}[htb!]
\centering
\includegraphics[width=\columnwidth]{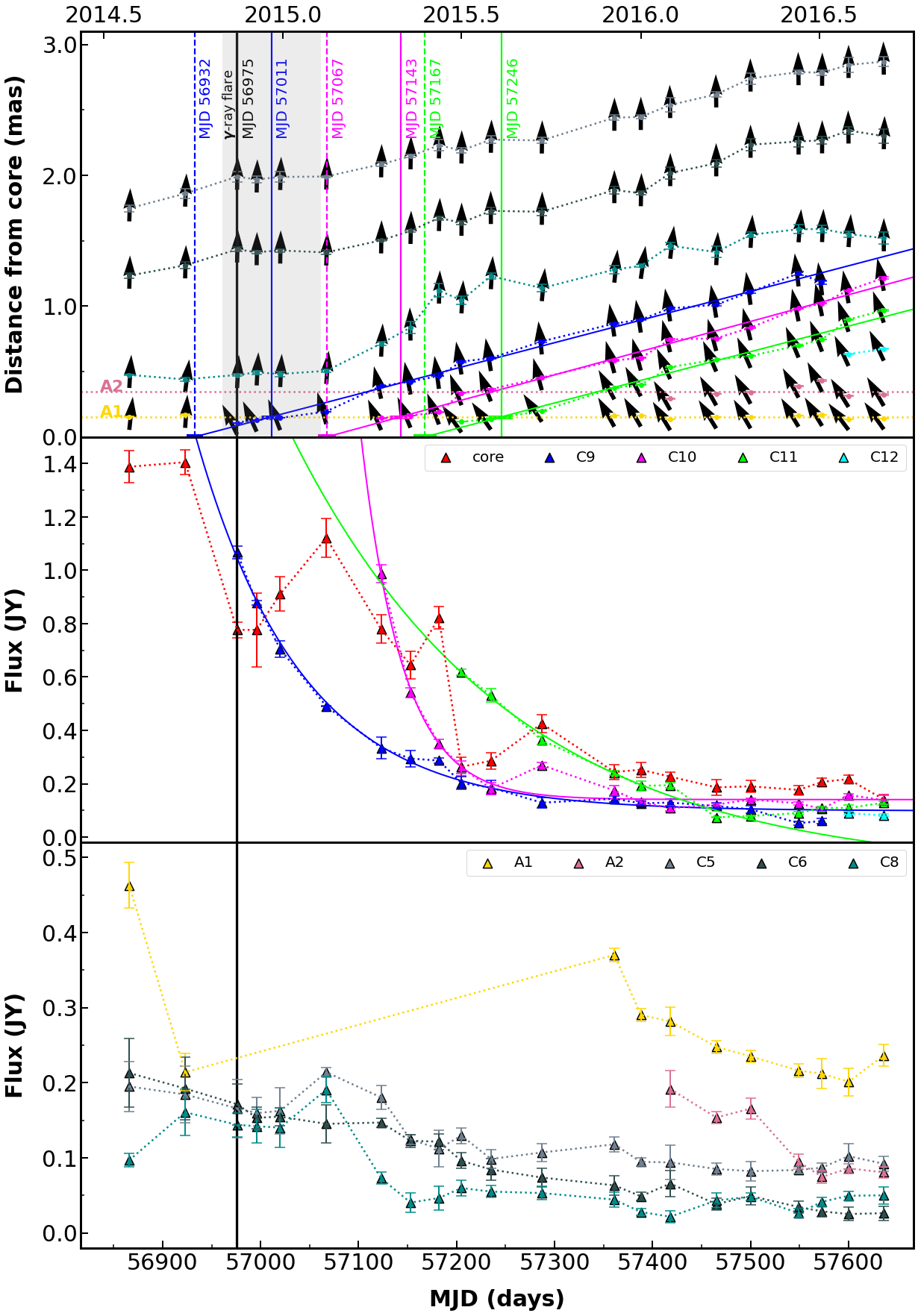}
\caption{The upper panel shows the radial distance of 43~GHz jet components relative to the core, which is shifted to the phase center at (0,0) position, from April 2015 to September 2016. The black arrows represent their position angles, and the solid lines represent their fitted trajectories. The colors correspond to the newly ejected components: C9 (blue), C10 (magenta), C11 (green), and C12 (cyan). The gray box highlights the interaction period between the C9 and A1 components. Horizontal yellow and pink lines indicate the mean positions of stationary components A1 and A2, respectively. The second panel presents the flux density of the core (red) and the newly ejected components, C9, C10, C11, and C12. The last panel shows the flux density of components with relatively low flux density, C5, C6, C8, C12, A1, and A2. The black vertical line at MJD 56975 represents the epoch of the $\gamma$-ray flare.
} \label{fig:fig6}
\vspace{5mm}
\end{figure}

For the iMOGABA data, the selected decaying periods are MJD~57107--57680 (11~epochs) at 22~GHz, MJD~57142--57623 (9~epochs) at 43~GHz, and MJD~57142--57719 (10~epochs) at 86~GHz.
For the VLBA 43~GHz, we selected the data from MJD~57124--57680 (15~epochs) for the decaying period.  
For the 15~GHz observations of MOJAVE, we used the data covering MJD~57189--57823 (8~epochs).
For the model-fitting results of the 43~GHz BU data, the selected epochs for the C9, C10, and C11 components are MJD~56976--57549 (16~epochs), MJD~57123--57573 (13~epochs), and MJD~57205--57465 (7~epochs), respectively.
In the second panel in Fig.~\ref{fig:fig6}, the solid lines represent the results of exponential fits to the flux decay of each component.

The best-fitting decay timescales obtained using the above method are summarized in Table~\ref{tab:table3}, yielding $404\pm130$~days at 15 GHz, $401\pm88$~days at 22~GHz, $227\pm62$~days at 43~GHz, and $150\pm42$ days at 86~GHz.
The timescales for the C9, C10, and C11 components are $108\pm6$, $43\pm4$, and $222\pm88$~days, respectively, as presented in Table~\ref{tab:table2}.

\subsection{Physical parameters of the jet components} \label{3.5}
We identified three newly emerging superluminal knots (C9, C10, and C11) in the source at 43~GHz. 
The estimated physical parameters and their corresponding uncertainties for these components are summarized in Table~\ref{tab:table2}.
By fitting a linear function to the angular distance from the core over time, we determined the proper motion, $\mu_{\mathrm{ang}}$ for each component. 
The corresponding apparent speed, $\beta_{\mathrm{app}}$ (in units of the speed of light), was found to be approximately 13.4 for all of the newly ejected components, with a typical uncertainty of $\sim 0.5$.
We assumed that the decay in the light curve is attributable to synchrotron cooling~\citep[e.g.,][and see Appendix~\ref{App1}]{lee2019jet}.
We regard the decay timescale $\tau$ as the synchrotron cooling timescale $\tau_{\mathrm{cool}}$.
Based on the assumption of the synchrotron cooling knots, we can use the cooling timescales $\tau_{\mathrm{cool}}$ to determine the variability Doppler factor $\delta_{\mathrm{var}}$ of the components as follows~\citep{Jorstad2005, jorstad2017kinematics}:
\begin{equation}
\label{eq3}
    \delta_{\mathrm{var}} = \frac{15.8sD_\text{L}}{\tau_{\mathrm{cool}}(1+z)},
\end{equation}
where $s$ is the angular size of the component in mas, taken as 1.8 times the FWHM for an optically thin spherical brightness distribution~\citep{1995ASPC...82..267P}; $D_\text{L}$ is the luminosity distance to the source in Gpc; and $\tau_{\mathrm{cool}}$ is the cooling timescale in years.
The variability Doppler factors of the superluminal knots range from $4\pm1.8$ to $22\pm2.4$.

In addition to the variability Doppler factor obtained from the equation \ref{eq3}, we also sought to estimate the Doppler factor using the brightness temperature $T_{\text{b}}$ of the components, as follows:
\begin{equation}
\label{eq4}
    T_{\text{b}} = 1.22 \times10^{12}\frac{S}{s^2 \nu^2}(1+z) ~\mathrm{K},
\end{equation}
where $S$ is the flux density at the peak in Jy, $s$ is the angular size in mas, $\nu$ is the observing frequency in GHz (i.e., 43~GHz), and $z$ is the redshift of the source ($z=0.435$). 
The Doppler factor $\delta$ is then estimated as the ratio of the maximum intrinsic brightness temperature $T_\text{int}$ to the observed brightness temperature, as follows~\citep{lahteenmaki1999total}:
\begin{equation}
\label{eq5}
\delta =\frac{T_\text{b}}{T_\text{int}}.
\end{equation}
We assume that $T_\text{int} \approx T_\text{eq}$ and adopt the equipartition value of $5 \times 10^{10}$~K suggested in \cite{readhead1994equipartition}.
The brightness temperature varies between  $(0.57\pm0.02)\times10^{11}$~K and $ (0.99\pm0.04)\times10^{11}$~K, with the Doppler factors of the superluminal knots ranging from $1.1\pm0.1$ to $2.0\pm0.1$.
The viewing angles $\Theta$ are calculated with the apparent speeds and Doppler factors from equation~\ref{eq5} as follows:

\begin{equation}
\label{eq6}
\Theta = \arctan\left(\frac{2\beta_{\mathrm{app}}}{\beta_{\mathrm{app}}^2+\delta^2-1}\right).
\end{equation}
The viewing angles between the jet axis of the superluminal knots at 43~GHz range from  $8.2\pm0.2^\circ$ to $8.6\pm0.4^\circ$.

The projected apparent speed $\beta_{\mathrm{app}}$ is related to the intrinsic speed $\beta$ and the viewing angle $\Theta$ with respect to the line of sight as follows:
\begin{equation}
\label{eq7}
    \beta = \frac{\beta_{\mathrm{app}}}{\mathrm{sin}\Theta+\beta_{\mathrm{app}} \mathrm{cos}\Theta}.
\end{equation}
The intrinsic speeds approach the speed of light for all components ($\beta \approx 0.999$), yielding Lorentz factors of $\Gamma \sim 45\pm3$--$78\pm7$.

\begin{table}[ht]
\caption{Cooling timescales and magnetic field strengths of PKS~1222+216 at multi-frequencies\label{tab:table3}}
\centering
\setlength{\tabcolsep}{4.4pt} 
\renewcommand{\arraystretch}{1.1} 

\begin{tabular}{lccc}
    \hline
    \hline
\rule{0in}{3ex} (1) & (2) & (3) & (4) \\ 
Data & Period  & $\tau_\nu \approx \tau_{\mathrm{cool}}$ & $B_\nu $ 
\\ & (MJD) & (days) & (mG) 
\\
\hline
15~GHz MOJAVE \rule{0pt}{2.5ex}  & 57189-57823 & $404\pm130$ & $19\pm3$ \\
22~GHz iMOGABA  & 57107-57680  & $401\pm88$ & $19\pm2$ \\
43~GHz iMOGABA+BU & 57123-57636 & $227\pm62$ & $26\pm4$ \\
86~GHz iMOGABA & 57142-57719 & $150\pm42$ & $32\pm5$ \\    

\hline
\end{tabular}
 
\tablefoot{Column designations: (1) data used, (2) observation period used for fitting, (3) cooling timescale, and (4) magnetic field strength.}
\end{table}

\begin{table*}
\centering
\caption{Physical parameters in jet components}
\label{tab:table2}
\small
\renewcommand{\tabcolsep}{2.5mm}
\renewcommand{\arraystretch}{1.75}
\begin{tabular}{cccccccccc}

    \hline
    \hline
    \rule{0in}{1ex} (1) & (2) & (3) & (4) & (5) & (6) & (7) & (8) & (9) & (10) \\
    \rule{0in}{1ex}
Knot & $\mu_{\text{ang}}$ & $\beta_{\text{app}}$ & $\Gamma$ & $\tau_{\text{cool}}$ & $\delta_\text{var}$ & $T_\text{b}$
 & $\delta$ & $\Theta$  & $B$ \\ & (mas/year) & $(c)$ & & (days) & & ($10^{11}$~K) & & $(^\circ)$ & (mG)
 \\ [1ex]
\hline
C9 & $0.72\pm0.02$ & $13.22\pm0.31$ & $45\pm3$ & $108\pm6$ & $9.1\pm0.5$ & $0.99\pm0.04$ & $2.0\pm0.1$ & $8.46\pm0.19$  &  $83\pm3$ \\

C10 & $0.74\pm0.02$ & $13.77\pm0.41$ & $50\pm4$ & $43\pm4$ &  $22\pm2.4$ &  $0.97\pm0.06$  & $2.0\pm0.1$  &  $8.15\pm0.23$ &  $134\pm8$ \\

C11 & $0.71\pm0.03$ & $13.21\pm0.58$ &  $78\pm7$ & $222\pm88$ &  $4.4\pm1.8$  &  $0.57\pm0.02$ &  $1.1\pm0.1$ &  $8.59\pm0.37$ &  $77\pm15$ \\
\hline
\end{tabular}
\vspace{-1mm} 

\tablefoot{Column designations: (1) component name, (2) angular speed, (3) apparent speed, (4) Lorentz factor, (5) cooling timescale, (6) variability Doppler factor, (7) brightness temperature, (8) Doppler factor, (9) viewing angle, and (10) magnetic field strength.}
\end{table*}

\subsection{Magnetic field strengths} \label{3.6}
When a gyrating electron radiates power, the synchrotron cooling time can be defined via $\tau_{\mathrm{cool}}=E/\dot{E}$, with the electron energy $E=\gamma_\mathrm{e} m_\mathrm{e}c^2$, where $\gamma_\mathrm{e}$ is the electron Lorentz factor, $m_\mathrm{e}$ is the electron mass, $c$ is the speed of light, and $\dot{E} $ is d$E$/d$t$ with time $t$.
Using the Larmor formula and assuming an isotropic distribution of the pitch angle, $\dot{E}$ can be expressed in CGS units as $\langle P \rangle=\frac{4}{3}\sigma_TcU_B\gamma_\mathrm{e}^2\beta_\mathrm{e}^2$, where $\sigma_T$ is the Thomson scattering cross section, $U_B$ is the magnetic energy density with $B^2/8\pi$ and $B$ is the magnetic field strength~\citep{rybicki1991radiative}.
After calculating all constants in SI units, we can express the cooling time equation for the magnetic field strength in the observed frame as follows:
\begin{equation}
\label{eq8}
    B = \sqrt{7.74\frac{1+z}{\delta\gamma_\mathrm{e}\tau_{cool}}}.
\end{equation}
Assuming that the dominant synchrotron power is emitted at the peak frequency $\nu_{\mathrm{syn}}$ of the synchrotron spectrum and that the IC spectrum is dominated by the synchrotron self Compton (SSC), the Lorentz factor of the electrons dominantly contributing to the SSC can be estimated as follows \citep{beckmann2012active}:
\begin{equation}
\label{eq9}
    \gamma_\mathrm{e} = \sqrt{\frac{3\nu_{\mathrm{IC}}}{4\nu_{\mathrm{syn}}}},
\end{equation}
where $\nu_\mathrm{IC}$ is the peak frequency of the IC spectrum.
Referring to the spectral analysis results for the 2014 flare of PKS~1222+216 as reported in \cite{ezhikode2022long} and \cite{adams2022variability}, the peak frequencies of the synchrotron and IC humps during the flaring period differ by approximately eight orders of magnitude~($\nu_\text{IC}/\nu_\text{syn}\approx10^{8}$), yielding an estimated electron Lorentz factor $\gamma_\mathrm{e}$ of 8660.
This value is close to a value located at the upper end of the interval typically covered by FSRQs.
Furthermore, the magnetic field strength was determined to be $19\pm3~\mathrm{mG}$ at 15~GHz, $19\pm2~\mathrm{mG}$ at 22~GHz, $26\pm4~\mathrm{mG}$ at 43~GHz, and $32\pm5~\mathrm{mG}$ at 86~GHz, as presented in Table~\ref{tab:table3}. These were calculated based on the redshift $z=0.435$, the cooling timescales in Section \ref{3.4}, and the Doppler factors ($\delta_{22}=10.7\pm0.4$, $\delta_{43}=9.6\pm0.6$) derived from the inverse-variance weighted mean of jet components in \citet{lee2019jet}.
The magnetic field strength values of C9, C10, and C11 were estimated to be $83\pm3$, $134\pm8$, and $77\pm15~\mathrm{mG}$ using the timescales determined in Section \ref{3.4} and Doppler factor in Section \ref{3.5}, as summarized in Table~\ref{tab:table2}.
As a result, we found that the magnetic field strength decreases as a function of the frequency and that the region of the C10 component at 43~GHz has the most significant magnetic field strength.

\section{Discussion} \label{4}

\subsection{Connection between the $\boldsymbol{\gamma}$-ray flare and radio components}\label{4.1}
Several investigations have indicated a correlation between the emergence of new jet components and flaring activity~\citep{jorstad2001multiepoch,2012A&A...537A..70S, lico2022new,pedrosa2024flaring}.
Motivated by this, we determined the times at which the newly emerged components C9, C10, and C11 detached from the core by fitting a linear function to their measured separations.
The ejection times from the core for the components are as follows: MJD~$56932\pm8$ for C9, MJD~$57067\pm9$ for C10, and MJD~$57167\pm12$ for C11.
Additionally, we determined the interaction times with the stationary component A1 based on the trajectory fitting results, defined as the time at which the moving components passed through the centroid position of A1, located at $0.15\pm0.01$~mas from the core: MJD~$57011\pm11$ for C9, MJD~$57143\pm11$ for C10, and MJD~$57246\pm14$ for C11.
In the first panel in Fig.~\ref{fig:fig6}, the vertical dashed lines indicate the epochs when each component was ejected from the core, while the solid vertical lines mark the estimated interaction times between the moving components and A1. 
The emergence of these new components from the core is not temporally coincident with the $\gamma$-ray flare observed at MJD~56974.6--56977.8. 
For instance, the flare occurred 43~days after the emergence of C9 and 92~days before the emergence of C10. 
However, given that the interaction time between C9 and the stationary component A1 is only 36 days after the $\gamma$-ray flare, we can consider the possibility of interaction between the A1 and C9 components, affecting the $\gamma$-ray flare.

We estimated the duration of the interaction between C9 and A1 based on the angular sizes of A1 and C9, yielding an interaction period from MJD~56961 to MJD~57061 (as indicated by the gray box in Fig.~\ref{fig:fig6}). The $\gamma$-ray peak time (MJD~56975) coincides with this interaction period, suggesting that the $\gamma$-ray flare may be related to the interaction. 
This exhibits characteristics consistent with prior studies of interactions between the core or stationary components and moving components induced by a conical standing shock~\citep[e.g.,][]{jorstad2014jet, li2024multi}. 
Therefore, we can suggest that the new jet component C9 passes through the stationary component A1, as illustrated in Figure 7 of \cite{li2024multi}, accelerating synchrotron particles for IC scattering, which up-scatters low-energy photons (e.g., BLR photons) into $\gamma$-ray photons.

Unlike the situation with C9, a $\gamma$-ray flare does not occur when C10 and C11 traverse A1 (e.g., after MJD~57123).
When C10 and C11 emerged, the synchrotron particles may not have been accelerated for IC scattering (e.g., optimal shock environments are not formed) as were those for the interaction between C9 and A1.
As shown in Fig.~\ref{fig:fig5}, the inner region of the jet appears to bend eastward starting from the third epoch, when component C9 emerges (Table~\ref{tab:table1}; position angle changes from approximately $-4^\circ$ for A1 to $26^\circ$ for C9).  
In contrast, no significant change in the jet direction is observed during the ejection of components C10 and C11.
While this suggests a possible connection between the $\gamma$-ray flare and the jet bending, further investigation (e.g., a high-angular-resolution polarization study) is required to confirm this relationship in the future.

Another possibility involves the acceleration for component C10. Previous studies (e.g., \cite{2009ApJ...706.1253H,2012A&A...537A..70S,2018ApJ...864...67S}) have shown evident acceleration of jet components associated with standing shocks.
\cite{2012A&A...537A..70S}, for example, reported that 43~GHz VLBI components in 3C 345 accelerate over $\sim 23$~pc after passing the core, coinciding with $\gamma$-ray emission. Analogously, the $\gamma$-ray flare could arise during the passage of C10 through the core, followed by a $\sim90$~day period of acceleration toward A1, after which the component maintained a constant speed as identified in our kinematic analysis. Alternatively, the $\gamma$-ray flare could occur further upstream, within the BLR or dusty torus, followed by the physical acceleration of the plasma, which eventually emerges as component C10. However, as described in the following Section~\ref{4.2} regarding the estimated locations of the 43~GHz core and dusty torus, such upstream regions are expected to be highly opaque to high energy photons. The dense external radiation fields from the dusty torus would lead to significant $\gamma \gamma$ pair production, effectively absorbing the $\gamma$-rays before they can escape~\citep{2003APh....18..377D}. Therefore, we propose the interaction between C9 and A1 as the most plausible scenario for the 2014 $\gamma$-ray flare, given that its downstream location provides a more transparent environment for $\gamma$-rays, in addition to aligning with both the shorter time lag between the flare and C9’s separation and the observed jet bending.

\subsection{Estimation of the $\boldsymbol{\gamma}$-ray flare location and the seed photon}\label{4.2}
In AGN research, $\gamma$-ray emission is likely to be generated through the IC scattering of seed photons by relativistic particles.
The seed photons may be synchrotron photons originating within the jet itself for synchrotron self-Compton (SSC) scattering or may come from external sources such as BLR, dusty torus, and cosmic microwave background (CMB) radiation.

To investigate the origin of the seed photons for the IC process during the $\gamma$-ray flare, we constrained the location of the $\gamma$-ray flare caused by the interaction between the moving jet component C9 and the stationary component A1.
The flare location can be estimated by determining the distance of A1 from the central engine in parsecs. 
Following the model outlined by \cite{lobanov1998ultracompact}, the position of the VLBI core is determined under the assumption that the core is the optical depth $\tau=1$ surface.
The distance of the core from the central engine, $r_{\text{core}}$, depends on the observing frequency $\nu$, as described by $r_{\text{core}}\propto\nu^{-1/k_\text{r}}$, where $k_\text{r}$ is a parameter depending on the magnetic field and particle density distribution along the jet.
The absolute distance of the central engine can be estimated as follows:
\begin{equation}
\label{eq10}
r_{\text{core}}(\nu) = \Omega_{r\nu} \left[ \nu^{\frac{1}{k_\text{r}}} \sin \Theta \right]^{-1},
\end{equation}
where $\Omega_{r\nu}$ is the core shift measure obtained from the angular offset $\Delta r_{\text{mas}}$ in mas between core positions at two frequencies, $\nu_1$ and $\nu_2$ ($\nu_1 < \nu_2$):
\begin{equation}
\label{eq11}
\Omega_{r\nu} = 4.85 \times 10^{-9} \frac{\Delta r_{\text{mas}} D_\text{L} }{(1 + z)^2}\cdot \frac{\nu_1^{1/k_\text{r}} \nu_2^{1/k_\text{r}}}{\nu_2^{1/k_\text{r}} - \nu_1^{1/k_\text{r}}}
\end{equation}
It is expected for ideal measurements that $\Omega_{r\nu}$ is constant for all frequency pairs.
We used a core shift measurement of 0.178~mas between 8.1~GHz and 15.4~GHz, as reported in \cite{ezhikode2022long}, yielding $\Omega_{r8,15}=7.08$~pc~GHz.
By adopting a viewing angle of $\Theta=5.3^\circ$ \citep{pushkarev2017mojave} and assuming $k_\text{r} \approx 1$ (i.e., equipartition between the energy density of the jet particles and the magnetic field) in Equation \ref{eq10}, we derived the distance of the core at 43~GHz, $r_\text{core}(\mathrm{43~GHz})=1.765$~pc.
Given the averaged angular position of A1 at $0.15\pm0.01$~mas and a luminosity distance of 2.4~Gpc for the source, the projected distance of A1 from the core is estimated to be $r_\text{A1} = 9.35 \pm 0.81$~pc, considering the aforementioned jet viewing angle.
Consequently, the absolute distance of A1 from the central engine is determined to be $r_\text{A1}=11.11\pm0.8$~pc.
The location of the $\gamma$-ray flare can also be estimated using the observed time delay, $\Delta t_\text{obs}$, between the $\gamma$-ray peak time and the moment when C9 passes through A1.
Considering cosmological effects, the distance between the A1 component and the $\gamma$-ray emission region in the source frame can be expressed as follows:
\begin{equation}\label{eq12}
\Delta r = \frac{\beta_\text{app}c\Delta t_\text{obs}}{(1+z)\sin\Theta}
\end{equation}
Subsequently, the location of the $\gamma$-ray flare can be determined as $r_\gamma = r_\text{A1}-\Delta r_\text{C9, A1}$.
As the $\gamma$-ray peak time is used as MJD~56975 and the interaction time as MJD~57011, the location of the $\gamma$-ray flare is estimated to be approximately $r_\gamma = 9.2\pm1.0$~pc.
Taking into account the position and size of A1, the location of C9 at the moment when it first passes through A1 is estimated to be $10.95\pm0.82$~pc.
Given the uncertainty range of the $\gamma$-ray flare location and the position where C9 intersects with A1, it is plausible to conclude that the $\gamma$-ray flare occurred as C9 passed through A1.

Furthermore, the outer radius of the BLR and the radius of the dusty torus reported by \cite{ezhikode2022long} are $R_\text{BLRout} = 0.07$~pc and $R_\text{DT} = 2.27$~pc, respectively. 
The estimated location of the 43~GHz core ($\sim1.8$~pc) lies within this dusty torus region. As discussed in Section~\ref{4.1}, if the $\gamma$-ray flare had originated at the core, the high energy emission would likely have been attenuated by absorption within the dusty torus.
Moreover, the $\gamma$-ray flare location estimated from downstream jet components is significantly offset from the BLR and dusty torus by 10.5~pc and 8.35~pc, respectively.
Consequently, the seed photons participating in the IC mechanism for the $\gamma$-ray flare are less likely to have originated from the BLR or dusty torus and are most likely to have originated from the jet itself or from external CMB radiation.
Alternatively, as proposed by \cite{2010ApJ...710L.126M}, a slow sheath surrounding the jet spine could serve as a source of infrared seed photons produced by moving knots or standing shocks.
This feasibility of high energy emission far from the central engine is consistent with the scenario where the 2010 $\gamma$-ray flare of PKS~1222+216 originated outside the BLR due to jet recollimation~\citep{tavecchio}.

\begin{figure}[t]
\centering
\includegraphics[width=0.8\linewidth]{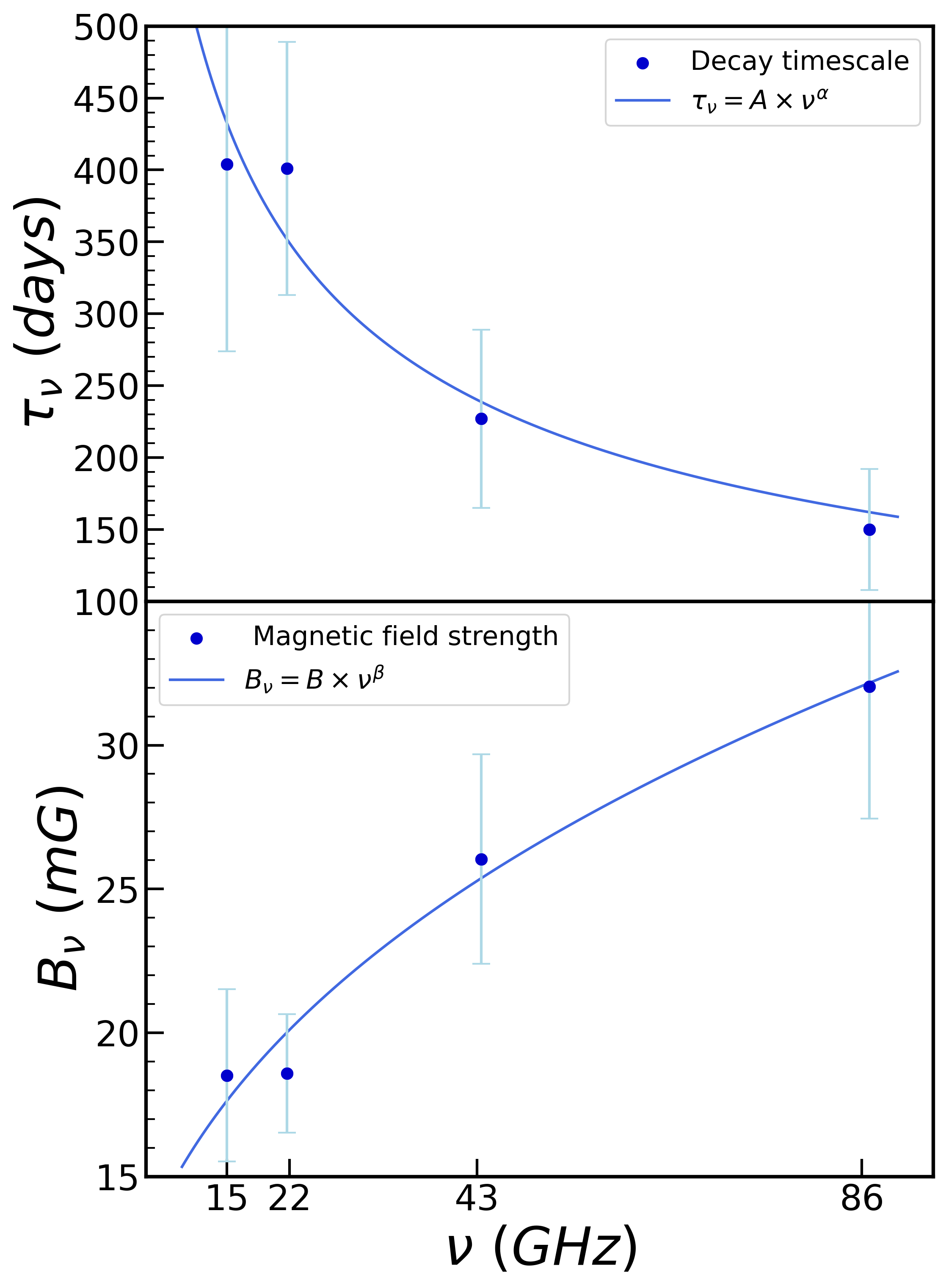}
\caption{The first panel displays the power-law fit of the cooling timescales as a function of frequency ($\tau_{\nu} = A \times \nu^{\alpha}$), while the second panel shows the power-law fit of the magnetic field strength as a function of frequency ($B_{\nu} = B \times \nu^{\beta}$). The best-fit power-law indices are $\alpha=-0.56\pm0.13$ and $\beta=0.34\pm0.04$. }\label{fig:fig7}
\vspace{5mm} 
\end{figure}

\subsection{Magnetic field strengths at multi-wavelengths}\label{4.3} 

In Section \ref{3.6}, we determined the magnetic field strengths to be $B\sim$20--30~mG in the radio core regions at frequencies of 15--86~GHz.
This result is consistent with the typical blazar magnetic field of $B\sim$10--100~mG~\citep{lewis2016time}.
We found that the magnetic field strength increases as the frequency increases.

The multi-wavelength magnetic field strengths and cooling timescales were fitted to a power-law function of the frequency ($\tau_\nu\propto\nu^\alpha$ and $B_\nu\propto\nu^\beta$), yielding the best-fit solution of $\alpha=-0.56\pm0.13$ and $\beta=0.34\pm0.04$, as shown in Fig.~\ref{fig:fig7}.
These results indicate that $\tau_{\mathrm{cool}}$ decreases exponentially and the magnetic field strengths increase as a function of frequency for an optically thin plasma.

The relationship between the cooling timescale and frequency ($\tau_\nu\propto \nu ^{-0.56 \pm 0.13}$) is consistent with the theoretical prediction ($\tau_\nu\propto \nu ^{-0.5}$) of an optically thin synchrotron plasma with critical frequency $\nu_\text{c}(\propto\gamma^3 \omega_\text{B})$, an electron angular frequency $\omega_\text{B}$ ($\propto \gamma^{-1}$), and cooling timescale $\tau(\propto \gamma^{-1})$ (see Equation \ref{eq8}).

The magnetic field strength is proportional to the distance from the central engine as $B\propto r^{-m}$, and the distance is proportional to the frequency as $r\propto \nu^{-1/k_\text{r}}$~(\cite{blandford1979relativistic}, \cite{o2010magnetic}), yielding $B\propto \nu^{m/k_\text{r}}$.
Assuming equipartition between the energy densities of jet particles and magnetic field energy densities with $k_\text{r} = 1$, the magnetic field strength $ B $ is expected to follow a radial profile proportional to $r^{-1} $ ($ B \propto r^{-1} $) following \cite{konigl1981relativistic}. 
However, the observational data for PKS~1222+216 indicate that the magnetic field strength in the jet evolves as $B \propto \nu^{0.34\pm0.04} $, indicating the quantity $m/k_\text{r}\approx0.34$.
This implies that the emission region in the relativistic jet of PKS~1222+216 may not be in an equipartition condition ($k_\text{r}\gtrsim 1$) or may have a magnetic field strength profile slowly decreasing along the jet ($m<1$).

\section{Conclusions} \label{5}

In this study, we investigated the long-term decline of multi-wavelength emissions of PKS~1222+216 using radio data from the iMOGABA program, MOJAVE, and the VLBA-BU-BLAZAR program spanning 2013--2020, combined with $\gamma$-ray data from \textit{Fermi}-LAT.
We found that the flux densities at radio frequencies of 15, 22, 43, and 86~GHz had decreased exponentially by 37\%--56\% over a year following a $\gamma$-ray flare in November 2014.
By attributing this decay to synchrotron cooling, we estimated key physical parameters of the source during the observation period.

Although previous studies have presented model fitting results for this source, we conducted our own independent Gaussian model fitting using VLBA 43~GHz data from July 2014 to September 2016. 
We identified ten jet components, including three newly ejected components (C9, C10, C11) and two stationary components (A1, A2).  
The cooling timescales for the new components were found to range from 43 to 222~days, with viewing angles confined within approximately eight degrees. 
Using the electron Lorentz factor derived under the assumption of SSC emission, we estimated magnetic field strengths for each component ranging from 77~mG to 134~mG.

We found that the interaction between C9 and A1 as the most plausible scenario for the 2014 $\gamma$-ray event, based on its temporal coincidence with the $\gamma$-ray flare and the physical constraint that the core region remains opaque to high energy photons due to absorption within the dusty torus.
To constrain the $\gamma$-ray emission site and the seed photon source for the IC process, we calculated the distance of component A1 from the central engine to be approximately $11.11\pm0.8$~pc, and the $\gamma$-ray flare location was estimated at $9.2\pm1.0$~pc. 
Given the absolute position and the size of component A1, taking into account the uncertainties, these results suggest that the flare occurred during the C9-A1 interaction, at a distance beyond the BLR and dusty torus region.
Therefore, the seed photons for the IC process likely originated from the jet itself, the CMB, or a surrounding sheath, rather than from the BLR or dusty torus. 
Our findings support a scenario where $\gamma$-ray emission occurs further downstream at stationary shocks, rather than close to the central engine.

Finally, we analyzed the magnetic field distribution based on optical depth variations, finding that the magnetic field scales as $B \propto r^{-0.34\pm0.04}$ along the jet.
This finding suggests that the emission region in the relativistic jet of PKS~1222+216 may deviate from the equipartition condition ($k_\text{r}\gtrsim 1$) or exhibit a slow decline in the magnetic field strength profile ($m<1$).
Our study demonstrates that analyzing cooling timescales offers a practical alternative method for determining magnetic field strengths in AGNs undergoing long-term synchrotron cooling.
\smallskip
\par\smallskip
{\fontsize{8}{9}\selectfont
\noindent\textit{Data availability.}
The KVN CLEAN images at 22, 43, and 86 GHz used in this study (shown in Figs.~\ref{fig:fig1} and \ref{fig:fig2}) are available in electronic form at the CDS via anonymous ftp to \url{http://cdsarc.u-strasbg.fr/} (130.79.128.5) or via \url{https://cdsarc.cds.unistra.fr}.
\par
}
\smallskip
\begin{acknowledgements}
We thank the anonymous reviewer for providing valuable comments that improved the manuscript.
We are grateful to the staff of the KVN who helped to operate the array and to correlate the data. The KVN is a facility operated by KASI (Korea Astronomy and Space Science Institute). The KVN observations and correlations are supported through the high-speed network connections among the KVN sites provided by KREONET (Korea Research Environment Open NETwork), which is managed and operated by KISTI (Korea Institute of Science and Technology Information). This work has made use of Fermi-LAT data supplied by \cite{abdollahi2023fermi}, \url{https://fermi.gsfc.nasa.gov/ssc/data/access/lat/LightCurveRepository}. This study makes use of VLBA data from the VLBA-BU Blazar Monitoring Program (BEAM-ME and VLBA-BUBLAZAR; \url{http://www.bu.edu/blazars/BEAM-ME.html}), funded by NASA through the Fermi Guest Investigator Program. The VLBA is an instrument of the National Radio Astronomy Observatory. The National Radio Astronomy Observatory is a facility of the National Science Foundation operated by Associated Universities, Inc. This work was supported by a National Research Foundation of Korea (NRF) grant funded by the Korean government (MIST) (2020R1A2C2009003, RS-2025-00562700).

\end{acknowledgements}

\bibliography{main_paper_bib}

\begin{appendix}
\section{Modelfitting parameters} 
We present the parameter values obtained from our independent model fitting in Table~\ref{tab:table1}.
These include the total flux density, peak flux density, distance from the core, component size with respect to the FWHM, and the position angle of the major axis, each presented with the corresponding uncertainty. 
The uncertainties of the parameters were estimated following the method described in Section~2.4 of \cite{Lee_2016}.
For the total flux density uncertainty, we adopted $\sigma_{\mathrm{tot}} = \sigma_{\mathrm{peak}} \,(S_{\mathrm{tot}}/S_{\mathrm{peak}})$.

\begin{table*}
\caption{Model fitting parameters\label{tab:table1}}
\small
\centering
\renewcommand{\tabcolsep}{7mm}
\renewcommand{\arraystretch}{1.1}
\begin{tabular}{@{}lcccccc@{}}
\hline
\hline

\rule{0in}{2ex} (1) & (2) & (3) & (4) & (5) & (6) & (7)\\
\rule{0in}{1ex}
Knot & MJD & $\mathrm{S}_{tot}~\pm~\sigma_{tot}$ & $\mathrm{S}_{peak}~\pm~\sigma_{peak}$ & $d~\pm~\sigma_d$  &$r~\pm~\sigma_r$ & $\theta~\pm~\sigma_{\theta}$\\
& (days) & (Jy) & (Jy/beam) & (mas) & (mas) & ($^\circ$) \\
 \hline

 Core & 56866 & $1.39~\pm~0.04$ & $1.33~\pm~0.04$ & $0.05~\pm~0.00$ & $0.0$ & $\ldots$ \\
		 & 56923 & $1.41~\pm~0.03$ & $1.32~\pm~0.03$ & $0.06~\pm~0.00$ & $0.0$ & $\ldots$ \\
		 & 56976 & $0.78~\pm~0.02$ & $0.76~\pm~0.02$ & $0.04~\pm~0.00$ & $0.0$ & $\ldots$ \\
		 & 56996 & $0.78~\pm~0.10$ & $0.73~\pm~0.09$ & $0.05~\pm~0.01$ & $0.0$ & $\ldots$ \\
		 & 57020 & $0.91~\pm~0.05$ & $0.83~\pm~0.04$ & $0.07~\pm~0.00$ & $0.0$ & $\ldots$ \\
		 & 57067 & $1.12~\pm~0.05$ & $1.10~\pm~0.05$ & $0.05~\pm~0.00$ & $0.0$ & $\ldots$ \\
		 & 57123 & $0.78~\pm~0.04$ & $0.75~\pm~0.04$ & $0.05~\pm~0.00$ & $0.0$ & $\ldots$ \\
		 & 57153 & $0.64~\pm~0.04$ & $0.60~\pm~0.03$ & $0.06~\pm~0.00$ & $0.0$ & $\ldots$ \\
		 & 57182 & $0.82~\pm~0.03$ & $0.78~\pm~0.03$ & $0.06~\pm~0.00$ & $0.0$ & $\ldots$ \\
		 & 57205 & $0.26~\pm~0.03$ & $0.26~\pm~0.02$ & $0.04~\pm~0.00$ & $0.0$ & $\ldots$ \\
		 & 57235 & $0.29~\pm~0.02$ & $0.28~\pm~0.02$ & $0.01~\pm~0.00$ & $0.0$ & $\ldots$ \\
		 & 57287 & $0.43~\pm~0.02$ & $0.39~\pm~0.02$ & $0.07~\pm~0.00$ & $0.0$ & $\ldots$ \\
		 & 57361 & $0.24~\pm~0.02$ & $0.23~\pm~0.02$ & $0.05~\pm~0.00$ & $0.0$ & $\ldots$ \\
		 & 57388 & $0.25~\pm~0.02$ & $0.25~\pm~0.02$ & $0.05~\pm~0.00$ & $0.0$ & $\ldots$ \\
		 & 57418 & $0.23~\pm~0.01$ & $0.23~\pm~0.01$ & $0.01~\pm~0.00$ & $0.0$ & $\ldots$ \\
		 & 57465 & $0.19~\pm~0.02$ & $0.18~\pm~0.02$ & $0.01~\pm~0.00$ & $0.0$ & $\ldots$ \\
		 & 57500 & $0.19~\pm~0.02$ & $0.19~\pm~0.01$ & $0.04~\pm~0.00$ & $0.0$ & $\ldots$ \\
		 & 57549 & $0.18~\pm~0.01$ & $0.17~\pm~0.01$ & $0.04~\pm~0.00$ & $0.0$ & $\ldots$ \\
		 & 57573 & $0.21~\pm~0.01$ & $0.20~\pm~0.01$ & $0.07~\pm~0.00$ & $0.0$ & $\ldots$ \\
		 & 57600 & $0.22~\pm~0.01$ & $0.21~\pm~0.01$ & $0.04~\pm~0.00$ & $0.0$ & $\ldots$ \\
		 & 57636 & $0.14~\pm~0.01$ & $0.13~\pm~0.01$ & $0.06~\pm~0.01$ & $0.0$ & $\ldots$ \\

   A1 & 56866 & $0.46~\pm~0.02$ & $0.40~\pm~0.02$ & $0.09~\pm~0.00$ & $0.15~\pm~0.00$ & $-6.12~\pm~0.01$ \\
		 & 56923 & $0.21~\pm~0.02$ & $0.19~\pm~0.02$ & $0.09~\pm~0.01$ & $0.17~\pm~0.00$ & $-3.95~\pm~0.02$ \\
		 & 57361 & $0.37~\pm~0.01$ & $0.35~\pm~0.01$ & $0.07~\pm~0.00$ & $0.17~\pm~0.00$ & $29.10~\pm~0.00$ \\
		 & 57388 & $0.29~\pm~0.01$ & $0.28~\pm~0.01$ & $0.06~\pm~0.00$ & $0.17~\pm~0.00$ & $30.02~\pm~0.00$ \\
		 & 57418 & $0.28~\pm~0.01$ & $0.27~\pm~0.01$ & $0.03~\pm~0.00$ & $0.14~\pm~0.00$ & $31.47~\pm~0.01$ \\
		 & 57465 & $0.25~\pm~0.01$ & $0.24~\pm~0.01$ & $0.03~\pm~0.00$ & $0.15~\pm~0.00$ & $31.81~\pm~0.00$ \\
		 & 57500 & $0.24~\pm~0.01$ & $0.22~\pm~0.01$ & $0.06~\pm~0.00$ & $0.15~\pm~0.00$ & $30.19~\pm~0.00$ \\
		 & 57549 & $0.22~\pm~0.01$ & $0.20~\pm~0.01$ & $0.06~\pm~0.00$ & $0.16~\pm~0.00$ & $31.37~\pm~0.01$ \\
		 & 57573 & $0.21~\pm~0.01$ & $0.20~\pm~0.01$ & $0.07~\pm~0.01$ & $0.17~\pm~0.00$ & $31.33~\pm~0.01$ \\
		 & 57600 & $0.20~\pm~0.01$ & $0.20~\pm~0.01$ & $0.00~\pm~0.00$ & $0.13~\pm~0.00$ & $36.93~\pm~0.00$ \\
		 & 57636 & $0.24~\pm~0.01$ & $0.22~\pm~0.01$ & $0.06~\pm~0.00$ & $0.14~\pm~0.00$ & $35.60~\pm~0.01$ \\

   A2 & 57418 & $0.19~\pm~0.02$ & $0.17~\pm~0.02$ & $0.09~\pm~0.01$ & $0.29~\pm~0.00$ & $29.31~\pm~0.01$ \\
		 & 57465 & $0.15~\pm~0.01$ & $0.14~\pm~0.01$ & $0.09~\pm~0.00$ & $0.33~\pm~0.00$ & $29.18~\pm~0.01$ \\
		 & 57500 & $0.17~\pm~0.01$ & $0.15~\pm~0.01$ & $0.11~\pm~0.01$ & $0.34~\pm~0.00$ & $28.23~\pm~0.01$ \\
		 & 57549 & $0.09~\pm~0.01$ & $0.08~\pm~0.01$ & $0.11~\pm~0.01$ & $0.39~\pm~0.00$ & $28.17~\pm~0.01$ \\
		 & 57573 & $0.07~\pm~0.01$ & $0.06~\pm~0.01$ & $0.13~\pm~0.01$ & $0.43~\pm~0.01$ & $25.87~\pm~0.01$ \\
		 & 57600 & $0.09~\pm~0.00$ & $0.08~\pm~0.00$ & $0.06~\pm~0.00$ & $0.31~\pm~0.00$ & $31.99~\pm~0.01$ \\
		 & 57636 & $0.08~\pm~0.01$ & $0.07~\pm~0.00$ & $0.12~\pm~0.01$ & $0.32~\pm~0.00$ & $30.23~\pm~0.01$ \\

         C5 & 56866 & $0.20~\pm~0.03$ & $0.08~\pm~0.01$ & $0.33~\pm~0.05$ & $1.75~\pm~0.03$ & $-1.74~\pm~0.01$ \\
		 & 56923 & $0.18~\pm~0.04$ & $0.07~\pm~0.01$ & $0.38~\pm~0.07$ & $1.86~\pm~0.04$ & $-2.35~\pm~0.02$ \\
		 & 56976 & $0.17~\pm~0.04$ & $0.08~\pm~0.02$ & $0.38~\pm~0.08$ & $1.99~\pm~0.04$ & $-1.02~\pm~0.02$ \\
		 & 56996 & $0.16~\pm~0.02$ & $0.06~\pm~0.01$ & $0.41~\pm~0.05$ & $1.97~\pm~0.03$ & $-1.20~\pm~0.01$ \\
		 & 57020 & $0.16~\pm~0.03$ & $0.06~\pm~0.01$ & $0.44~\pm~0.08$ & $1.99~\pm~0.04$ & $-1.67~\pm~0.02$ \\
		 & 57067 & $0.21~\pm~0.01$ & $0.11~\pm~0.00$ & $0.48~\pm~0.01$ & $1.99~\pm~0.01$ & $-1.48~\pm~0.00$ \\
		 & 57123 & $0.18~\pm~0.01$ & $0.08~\pm~0.01$ & $0.44~\pm~0.04$ & $2.08~\pm~0.02$ & $-1.00~\pm~0.01$ \\
		 & 57153 & $0.12~\pm~0.01$ & $0.05~\pm~0.00$ & $0.44~\pm~0.02$ & $2.15~\pm~0.01$ & $-0.93~\pm~0.01$ \\
		 & 57182 & $0.11~\pm~0.02$ & $0.04~\pm~0.01$ & $0.43~\pm~0.09$ & $2.23~\pm~0.04$ & $-1.46~\pm~0.02$ \\
		 & 57205 & $0.13~\pm~0.01$ & $0.04~\pm~0.00$ & $0.48~\pm~0.04$ & $2.19~\pm~0.02$ & $-0.51~\pm~0.01$ \\
		 & 57235 & $0.10~\pm~0.01$ & $0.03~\pm~0.00$ & $0.47~\pm~0.06$ & $2.27~\pm~0.03$ & $-1.17~\pm~0.01$ \\
		 & 57287 & $0.11~\pm~0.01$ & $0.04~\pm~0.00$ & $0.43~\pm~0.04$ & $2.27~\pm~0.02$ & $-0.69~\pm~0.01$ \\
		 & 57361 & $0.12~\pm~0.01$ & $0.05~\pm~0.00$ & $0.52~\pm~0.04$ & $2.44~\pm~0.02$ & $-0.27~\pm~0.01$ \\
		 & 57388 & $0.09~\pm~0.01$ & $0.05~\pm~0.00$ & $0.47~\pm~0.03$ & $2.45~\pm~0.01$ & $-0.15~\pm~0.01$ \\
		 & 57418 & $0.09~\pm~0.02$ & $0.02~\pm~0.01$ & $0.47~\pm~0.11$ & $2.54~\pm~0.05$ & $0.75~\pm~0.02$ \\
		 & 57465 & $0.09~\pm~0.01$ & $0.03~\pm~0.00$ & $0.51~\pm~0.04$ & $2.62~\pm~0.02$ & $0.76~\pm~0.01$ \\
		 & 57500 & $0.08~\pm~0.01$ & $0.02~\pm~0.00$ & $0.55~\pm~0.08$ & $2.74~\pm~0.04$ & $0.33~\pm~0.02$ \\
		 & 57549 & $0.08~\pm~0.00$ & $0.03~\pm~0.00$ & $0.60~\pm~0.03$ & $2.79~\pm~0.02$ & $0.63~\pm~0.01$ \\
		 & 57573 & $0.09~\pm~0.01$ & $0.03~\pm~0.00$ & $0.61~\pm~0.04$ & $2.78~\pm~0.02$ & $-0.42~\pm~0.01$ \\
		 & 57600 & $0.10~\pm~0.02$ & $0.03~\pm~0.00$ & $0.70~\pm~0.11$ & $2.84~\pm~0.06$ & $-0.09~\pm~0.02$ \\
		 & 57636 & $0.09~\pm~0.01$ & $0.02~\pm~0.00$ & $0.67~\pm~0.07$ & $2.87~\pm~0.04$ & $0.72~\pm~0.01$ \\

\end{tabular}
\end{table*}

\begin{table*}
\addtocounter{table}{-1} 
\caption{Continued}
\small
\centering
\renewcommand{\tabcolsep}{7mm}
\renewcommand{\arraystretch}{1.1}
\begin{tabular}{@{}lcccccc@{}}
\hline
\hline

\rule{0in}{2ex} (1) & (2) & (3) & (4) & (5) & (6) & (7)\\
\rule{0in}{1ex}
Knot & MJD & $\mathrm{S}_{tot}~\pm~\sigma_{tot}$  & $\mathrm{S}_{peak}~\pm~\sigma_{peak}$ & $d~\pm~\sigma_d$ & $r~\pm~\sigma_r$ & $\theta~\pm~\sigma_{\theta}$\\
& (days) & (Jy) & (Jy/beam) & (mas) & (mas) & ($^\circ$) \\
 \hline

C6 & 56866 & $0.21~\pm~0.04$ & $0.12~\pm~0.02$ & $0.24~\pm~0.04$ & $1.23~\pm~0.02$ & $-1.28~\pm~0.02$ \\
		 & 56923 & $0.19~\pm~0.04$ & $0.10~\pm~0.02$ & $0.26~\pm~0.05$ & $1.31~\pm~0.02$ & $-1.50~\pm~0.02$ \\
		 & 56976 & $0.17~\pm~0.02$ & $0.10~\pm~0.01$ & $0.26~\pm~0.04$ & $1.43~\pm~0.02$ & $-1.35~\pm~0.01$ \\
		 & 56996 & $0.15~\pm~0.01$ & $0.08~\pm~0.01$ & $0.28~\pm~0.02$ & $1.41~\pm~0.01$ & $-0.59~\pm~0.01$ \\
		 & 57020 & $0.15~\pm~0.01$ & $0.07~\pm~0.01$ & $0.32~\pm~0.02$ & $1.42~\pm~0.01$ & $-0.97~\pm~0.01$ \\
		 & 57067 & $0.15~\pm~0.02$ & $0.11~\pm~0.02$ & $0.25~\pm~0.03$ & $1.41~\pm~0.02$ & $-1.78~\pm~0.01$ \\
		 & 57123 & $0.15~\pm~0.01$ & $0.09~\pm~0.00$ & $0.29~\pm~0.01$ & $1.50~\pm~0.01$ & $-0.79~\pm~0.00$ \\
		 & 57153 & $0.12~\pm~0.01$ & $0.06~\pm~0.00$ & $0.32~\pm~0.02$ & $1.57~\pm~0.01$ & $-1.09~\pm~0.01$ \\
		 & 57182 & $0.12~\pm~0.01$ & $0.06~\pm~0.00$ & $0.34~\pm~0.03$ & $1.67~\pm~0.01$ & $-1.73~\pm~0.01$ \\
		 & 57205 & $0.10~\pm~0.01$ & $0.05~\pm~0.00$ & $0.31~\pm~0.03$ & $1.64~\pm~0.02$ & $-0.32~\pm~0.01$ \\
		 & 57235 & $0.08~\pm~0.01$ & $0.04~\pm~0.01$ & $0.35~\pm~0.05$ & $1.73~\pm~0.03$ & $0.31~\pm~0.02$ \\
		 & 57287 & $0.07~\pm~0.01$ & $0.03~\pm~0.01$ & $0.35~\pm~0.06$ & $1.72~\pm~0.03$ & $-0.69~\pm~0.02$ \\
		 & 57361 & $0.06~\pm~0.01$ & $0.03~\pm~0.01$ & $0.34~\pm~0.06$ & $1.89~\pm~0.03$ & $-0.08~\pm~0.02$ \\
		 & 57388 & $0.05~\pm~0.00$ & $0.03~\pm~0.00$ & $0.30~\pm~0.03$ & $1.86~\pm~0.01$ & $-1.20~\pm~0.01$ \\
		 & 57418 & $0.07~\pm~0.02$ & $0.02~\pm~0.01$ & $0.37~\pm~0.10$ & $2.02~\pm~0.05$ & $-0.25~\pm~0.02$ \\
		 & 57465 & $0.04~\pm~0.01$ & $0.02~\pm~0.00$ & $0.32~\pm~0.06$ & $2.09~\pm~0.03$ & $0.15~\pm~0.01$ \\
		 & 57500 & $0.05~\pm~0.01$ & $0.02~\pm~0.00$ & $0.41~\pm~0.09$ & $2.23~\pm~0.04$ & $0.01~\pm~0.02$ \\
		 & 57549 & $0.03~\pm~0.01$ & $0.02~\pm~0.00$ & $0.37~\pm~0.08$ & $2.26~\pm~0.04$ & $-0.28~\pm~0.02$ \\
		 & 57573 & $0.03~\pm~0.01$ & $0.02~\pm~0.00$ & $0.30~\pm~0.06$ & $2.27~\pm~0.03$ & $-0.23~\pm~0.01$ \\
		 & 57600 & $0.03~\pm~0.01$ & $0.02~\pm~0.01$ & $0.19~\pm~0.05$ & $2.34~\pm~0.03$ & $0.61~\pm~0.01$ \\
		 & 57636 & $0.03~\pm~0.01$ & $0.01~\pm~0.00$ & $0.36~\pm~0.11$ & $2.30~\pm~0.06$ & $-1.42~\pm~0.02$ \\

C8 & 56866 & $0.10~\pm~0.01$ & $0.08~\pm~0.01$ & $0.11~\pm~0.01$ & $0.47~\pm~0.00$ & $-3.53~\pm~0.01$ \\
		 & 56923 & $0.16~\pm~0.03$ & $0.11~\pm~0.02$ & $0.18~\pm~0.03$ & $0.44~\pm~0.01$ & $-7.06~\pm~0.03$ \\
		 & 56976 & $0.14~\pm~0.01$ & $0.11~\pm~0.01$ & $0.18~\pm~0.02$ & $0.47~\pm~0.01$ & $-3.12~\pm~0.02$ \\
		 & 56996 & $0.14~\pm~0.02$ & $0.09~\pm~0.01$ & $0.21~\pm~0.03$ & $0.49~\pm~0.01$ & $-2.98~\pm~0.03$ \\
		 & 57020 & $0.14~\pm~0.02$ & $0.08~\pm~0.01$ & $0.23~\pm~0.04$ & $0.48~\pm~0.02$ & $-2.78~\pm~0.04$ \\
		 & 57067 & $0.19~\pm~0.01$ & $0.12~\pm~0.01$ & $0.35~\pm~0.03$ & $0.50~\pm~0.01$ & $-4.15~\pm~0.03$ \\
		 & 57123 & $0.07~\pm~0.01$ & $0.05~\pm~0.00$ & $0.26~\pm~0.02$ & $0.71~\pm~0.01$ & $-2.78~\pm~0.02$ \\
		 & 57153 & $0.04~\pm~0.01$ & $0.02~\pm~0.01$ & $0.31~\pm~0.09$ & $0.84~\pm~0.04$ & $1.16~\pm~0.05$ \\
		 & 57182 & $0.05~\pm~0.01$ & $0.02~\pm~0.01$ & $0.34~\pm~0.10$ & $1.12~\pm~0.05$ & $-1.72~\pm~0.05$ \\
		 & 57205 & $0.06~\pm~0.01$ & $0.03~\pm~0.00$ & $0.38~\pm~0.06$ & $1.04~\pm~0.03$ & $-3.10~\pm~0.03$ \\
		 & 57235 & $0.06~\pm~0.01$ & $0.03~\pm~0.00$ & $0.32~\pm~0.04$ & $1.23~\pm~0.02$ & $-3.30~\pm~0.02$ \\
		 & 57287 & $0.05~\pm~0.01$ & $0.02~\pm~0.00$ & $0.37~\pm~0.06$ & $1.14~\pm~0.03$ & $-5.41~\pm~0.02$ \\
		 & 57361 & $0.04~\pm~0.01$ & $0.03~\pm~0.01$ & $0.32~\pm~0.06$ & $1.28~\pm~0.03$ & $-5.55~\pm~0.02$ \\
		 & 57388 & $0.03~\pm~0.00$ & $0.02~\pm~0.00$ & $0.17~\pm~0.02$ & $1.31~\pm~0.01$ & $-8.19~\pm~0.01$ \\
		 & 57418 & $0.02~\pm~0.01$ & $0.01~\pm~0.00$ & $0.16~\pm~0.05$ & $1.46~\pm~0.02$ & $-6.65~\pm~0.02$ \\
		 & 57465 & $0.04~\pm~0.01$ & $0.02~\pm~0.00$ & $0.42~\pm~0.09$ & $1.41~\pm~0.04$ & $-2.86~\pm~0.03$ \\
		 & 57500 & $0.05~\pm~0.00$ & $0.02~\pm~0.00$ & $0.46~\pm~0.04$ & $1.54~\pm~0.02$ & $-3.31~\pm~0.01$ \\
		 & 57549 & $0.03~\pm~0.01$ & $0.01~\pm~0.00$ & $0.35~\pm~0.07$ & $1.59~\pm~0.04$ & $-4.54~\pm~0.02$ \\
		 & 57573 & $0.04~\pm~0.01$ & $0.02~\pm~0.00$ & $0.44~\pm~0.06$ & $1.59~\pm~0.03$ & $-2.47~\pm~0.02$ \\
		 & 57600 & $0.05~\pm~0.01$ & $0.03~\pm~0.00$ & $0.42~\pm~0.05$ & $1.55~\pm~0.02$ & $-4.01~\pm~0.02$ \\
		 & 57636 & $0.05~\pm~0.01$ & $0.02~\pm~0.00$ & $0.44~\pm~0.09$ & $1.52~\pm~0.05$ & $-0.07~\pm~0.03$ \\

C9 & 56976 & $1.07~\pm~0.02$ & $1.02~\pm~0.02$ & $0.06~\pm~0.00$ & $0.11~\pm~0.00$ & $26.23~\pm~0.00$ \\
		 & 56996 & $0.88~\pm~0.01$ & $0.80~\pm~0.01$ & $0.06~\pm~0.00$ & $0.13~\pm~0.00$ & $24.29~\pm~0.00$ \\
		 & 57020 & $0.70~\pm~0.02$ & $0.64~\pm~0.02$ & $0.07~\pm~0.00$ & $0.14~\pm~0.00$ & $21.25~\pm~0.01$ \\
		 & 57067 & $0.49~\pm~0.00$ & $0.47~\pm~0.00$ & $0.08~\pm~0.00$ & $0.19~\pm~0.00$ & $15.60~\pm~0.00$ \\
		 & 57123 & $0.33~\pm~0.03$ & $0.31~\pm~0.03$ & $0.08~\pm~0.01$ & $0.39~\pm~0.00$ & $12.44~\pm~0.01$ \\
		 & 57153 & $0.29~\pm~0.02$ & $0.26~\pm~0.02$ & $0.10~\pm~0.01$ & $0.42~\pm~0.00$ & $10.27~\pm~0.01$ \\
		 & 57182 & $0.29~\pm~0.01$ & $0.24~\pm~0.01$ & $0.12~\pm~0.00$ & $0.47~\pm~0.00$ & $7.62~\pm~0.00$ \\
		 & 57205 & $0.20~\pm~0.02$ & $0.17~\pm~0.02$ & $0.13~\pm~0.01$ & $0.58~\pm~0.01$ & $9.28~\pm~0.01$ \\
		 & 57235 & $0.19~\pm~0.02$ & $0.14~\pm~0.02$ & $0.14~\pm~0.02$ & $0.60~\pm~0.01$ & $9.02~\pm~0.01$ \\
		 & 57287 & $0.13~\pm~0.01$ & $0.11~\pm~0.01$ & $0.13~\pm~0.01$ & $0.73~\pm~0.01$ & $5.88~\pm~0.01$ \\
		 & 57361 & $0.14~\pm~0.01$ & $0.11~\pm~0.01$ & $0.18~\pm~0.01$ & $0.87~\pm~0.01$ & $8.50~\pm~0.01$ \\
		 & 57388 & $0.13~\pm~0.01$ & $0.10~\pm~0.01$ & $0.17~\pm~0.01$ & $0.90~\pm~0.00$ & $8.43~\pm~0.01$ \\
		 & 57418 & $0.13~\pm~0.01$ & $0.08~\pm~0.01$ & $0.20~\pm~0.02$ & $0.98~\pm~0.01$ & $8.93~\pm~0.01$ \\
		 & 57465 & $0.12~\pm~0.01$ & $0.09~\pm~0.01$ & $0.19~\pm~0.02$ & $1.01~\pm~0.01$ & $9.25~\pm~0.01$ \\
		 & 57500 & $0.10~\pm~0.01$ & $0.07~\pm~0.01$ & $0.20~\pm~0.02$ & $1.11~\pm~0.01$ & $7.64~\pm~0.01$ \\
		 & 57549 & $0.05~\pm~0.01$ & $0.03~\pm~0.00$ & $0.26~\pm~0.03$ & $1.25~\pm~0.01$ & $6.63~\pm~0.01$ \\
		 & 57573 & $0.06~\pm~0.01$ & $0.04~\pm~0.01$ & $0.21~\pm~0.03$ & $1.18~\pm~0.01$ & $6.83~\pm~0.01$ \\             

\end{tabular}
\end{table*}

\begin{table*}
\addtocounter{table}{-1} 
\caption{Continued}
\small
\centering
\renewcommand{\tabcolsep}{7mm}
\renewcommand{\arraystretch}{1.1}
\begin{tabular}{@{}lcccccc@{}}
\hline
\hline

\rule{0in}{2ex} (1) & (2) & (3) & (4) & (5) & (6) & (7)\\
\rule{0in}{1ex}
Knot & MJD & $\mathrm{S}_{tot}~\pm~\sigma_{tot}$  & $\mathrm{S}_{peak}~\pm~\sigma_{peak}$ & $d~\pm~\sigma_d$ & $r~\pm~\sigma_r$  & $\theta~\pm~\sigma_{\theta}$\\
& (days) & (Jy) & (Jy/beam) & (mas) & (mas) & ($^\circ$) \\

 \hline

C10 & 57123 & $0.99~\pm~0.02$ & $0.94~\pm~0.02$ & $0.05~\pm~0.00$ & $0.14~\pm~0.00$ & $23.21~\pm~0.00$ \\
		 & 57153 & $0.54~\pm~0.01$ & $0.50~\pm~0.01$ & $0.07~\pm~0.00$ & $0.16~\pm~0.00$ & $23.05~\pm~0.01$ \\
		 & 57182 & $0.35~\pm~0.01$ & $0.33~\pm~0.01$ & $0.04~\pm~0.00$ & $0.19~\pm~0.00$ & $20.60~\pm~0.00$ \\
		 & 57205 & $0.26~\pm~0.02$ & $0.23~\pm~0.02$ & $0.09~\pm~0.01$ & $0.34~\pm~0.00$ & $20.20~\pm~0.01$ \\
		 & 57235 & $0.18~\pm~0.02$ & $0.17~\pm~0.02$ & $0.06~\pm~0.01$ & $0.36~\pm~0.00$ & $21.15~\pm~0.01$ \\
		 & 57287 & $0.27~\pm~0.01$ & $0.21~\pm~0.01$ & $0.14~\pm~0.00$ & $0.45~\pm~0.00$ & $14.85~\pm~0.00$ \\
		 & 57361 & $0.17~\pm~0.01$ & $0.14~\pm~0.01$ & $0.15~\pm~0.01$ & $0.58~\pm~0.01$ & $14.85~\pm~0.01$ \\
		 & 57388 & $0.13~\pm~0.01$ & $0.11~\pm~0.00$ & $0.14~\pm~0.01$ & $0.60~\pm~0.00$ & $14.57~\pm~0.00$ \\
		 & 57418 & $0.11~\pm~0.01$ & $0.08~\pm~0.01$ & $0.15~\pm~0.01$ & $0.75~\pm~0.01$ & $13.70~\pm~0.01$ \\
		 & 57465 & $0.12~\pm~0.00$ & $0.10~\pm~0.00$ & $0.14~\pm~0.00$ & $0.75~\pm~0.00$ & $14.87~\pm~0.00$ \\
		 & 57500 & $0.14~\pm~0.00$ & $0.11~\pm~0.00$ & $0.17~\pm~0.01$ & $0.84~\pm~0.00$ & $13.25~\pm~0.00$ \\
		 & 57549 & $0.13~\pm~0.00$ & $0.09~\pm~0.00$ & $0.18~\pm~0.01$ & $0.98~\pm~0.00$ & $11.40~\pm~0.00$ \\
		 & 57573 & $0.11~\pm~0.01$ & $0.08~\pm~0.00$ & $0.17~\pm~0.01$ & $1.02~\pm~0.00$ & $11.69~\pm~0.00$ \\
		 & 57600 & $0.16~\pm~0.00$ & $0.12~\pm~0.00$ & $0.20~\pm~0.01$ & $1.12~\pm~0.00$ & $9.42~\pm~0.00$ \\
		 & 57636 & $0.14~\pm~0.01$ & $0.10~\pm~0.01$ & $0.19~\pm~0.02$ & $1.21~\pm~0.01$ & $11.27~\pm~0.01$ \\

C11 & 57205 & $0.62~\pm~0.01$ & $0.59~\pm~0.01$ & $0.06~\pm~0.00$ & $0.12~\pm~0.00$ & $35.65~\pm~0.00$ \\
		 & 57235 & $0.53~\pm~0.02$ & $0.50~\pm~0.02$ & $0.06~\pm~0.00$ & $0.14~\pm~0.00$ & $34.83~\pm~0.01$ \\
		 & 57287 & $0.36~\pm~0.01$ & $0.33~\pm~0.01$ & $0.08~\pm~0.00$ & $0.20~\pm~0.00$ & $33.62~\pm~0.01$ \\
		 & 57361 & $0.24~\pm~0.01$ & $0.23~\pm~0.01$ & $0.07~\pm~0.00$ & $0.37~\pm~0.00$ & $28.31~\pm~0.00$ \\
		 & 57388 & $0.19~\pm~0.01$ & $0.18~\pm~0.01$ & $0.06~\pm~0.00$ & $0.40~\pm~0.00$ & $27.56~\pm~0.01$ \\
		 & 57418 & $0.19~\pm~0.01$ & $0.17~\pm~0.01$ & $0.08~\pm~0.00$ & $0.53~\pm~0.00$ & $23.54~\pm~0.00$ \\
		 & 57465 & $0.07~\pm~0.00$ & $0.07~\pm~0.00$ & $0.05~\pm~0.00$ & $0.59~\pm~0.00$ & $23.35~\pm~0.00$ \\
		 & 57500 & $0.08~\pm~0.00$ & $0.07~\pm~0.00$ & $0.09~\pm~0.00$ & $0.62~\pm~0.00$ & $23.50~\pm~0.00$ \\
		 & 57549 & $0.09~\pm~0.00$ & $0.08~\pm~0.00$ & $0.11~\pm~0.00$ & $0.70~\pm~0.00$ & $22.51~\pm~0.00$ \\
		 & 57573 & $0.11~\pm~0.00$ & $0.09~\pm~0.00$ & $0.15~\pm~0.01$ & $0.74~\pm~0.00$ & $21.30~\pm~0.00$ \\
		 & 57600 & $0.11~\pm~0.01$ & $0.10~\pm~0.01$ & $0.09~\pm~0.01$ & $0.90~\pm~0.00$ & $18.90~\pm~0.00$ \\
		 & 57636 & $0.13~\pm~0.01$ & $0.10~\pm~0.00$ & $0.16~\pm~0.01$ & $0.97~\pm~0.00$ & $18.94~\pm~0.00$ \\
         
C12 & 57600 & $0.09~\pm~0.00$ & $0.08~\pm~0.00$ & $0.08~\pm~0.00$ & $0.63~\pm~0.00$ & $24.63~\pm~0.00$ \\
		 & 57636 & $0.08~\pm~0.00$ & $0.06~\pm~0.00$ & $0.17~\pm~0.01$ & $0.67~\pm~0.00$ & $25.32~\pm~0.01$ \\

\hline                       
\end{tabular}
\vspace{0mm}
\tablefoot{Column designations: (1) component name, (2) MJD of the start of the observation, (3) total flux density and corresponding uncertainty, (4) peak flux density and corresponding uncertainty, (5) component size with respect to the FWHM and corresponding uncertainty, (6) distance from the core and corresponding uncertainty, and (7) position angle of the major axis and related uncertainty.}
\end{table*}

\section{Comparative analysis and identification of jet Components at 43~GHz} \label{App2}
Although \citet{Weaver_2022} presented the result of the Gaussian model fitting and component identifications for PKS 1222+216, we independently analyzed the 43 GHz BU data from mid-2014 to mid-2016, applying the criteria outlined in Section \ref{2.2} and \ref{3.3}.
We found fewer components in each epoch than \cite{Weaver_2022}.
For example, they detected eight components with the core, but we detected six components on MJD~57123.
We also detected C9 and C11 two epochs earlier than reported by \cite{Weaver_2022}.
When a new component is detected in the vicinity of the stationary A1 region ($r\approx0.14$~mas), it is necessary to determine whether the component is a reappearing A1 component or a newly emerged jet component.
In such cases, when a new component was detected at the typical position of A1 in the MJD~56976 epoch, we identified it as a newly emerged jet component (e.g., C9) based on its flux variation. 
In the previous epoch (MJD~56923), A1 had a flux density of 0.2~Jy, whereas the newly detected component exhibited a sudden increase to 1.1~Jy on MJD~56976. 
This abrupt flux enhancement indicates that the detected component is distinct from A1 and should be classified as a newly emerged jet component (C9).
In addition to the flux variation, we also considered the distance from the core and the position angle. 
In the first two epochs, the component closest to the core was identified as the stationary A1, while the following component was classified as C9 due to its abrupt change in position angle (see Fig.~\ref{fig:fig6}).
In the previous study, three components of B3, B5, and B6 were detected at distances greater than 1~mas from the core, whereas in our results, only two components, C5 and C6, were detected (i.e., non-detection of the B3 component).
In \cite{Weaver_2022}, the stationary A1 component was detected starting from the 14th epoch (MJD~57388).
However, in our results, A1 was revealed in the first and second epochs (MJD~56866 and 56923) and then reappeared in the 13th epoch (MJD~57361).
The stationary A2 component was also detected one epoch later (MJD~57418) in our result compared to \cite{Weaver_2022}. 
Nonetheless, within the range of uncertainty, the average distance from the core and the flux values of A1 and A2 were consistent with the previous result (see Table~\ref{tab:table4}).
We also found that the appearance epochs of C8 and C10 coincide with those reported in \cite{Weaver_2022} (MJD~56866 and MJD~57124).

\begin{table}[t!]
\caption{Comparison of the physical properties of stationary components with \cite{Weaver_2022}\label{tab:table4}}
\centering
\renewcommand{\tabcolsep}{1.5mm}
\begin{tabular}{ccccc}
    \hline
    \hline
\rule{0in}{3ex} (1)                          & (2)                                   & (3)                           & (4)        & (5)                        \\[1ex]
Knot                        & $\langle S \rangle$ - Weaver                      & $\langle S \rangle$                                  & $\langle r \rangle$ - Weaver & $\langle r \rangle$  \\

&  (Jy) & (Jy) & (mas) & (mas)  \\
\hline\\

A1    & $0.26\pm0.12$     & $0.27\pm0.08$   & $0.15\pm0.02$   & $0.15\pm0.01$  \\

A2    & $0.14\pm0.05$     & $0.12\pm0.05$   & $0.32\pm0.05$   & $0.35\pm0.05$  \\

\hline
\end{tabular}
\tablefoot{Column designations: (1) name of the stationary component, (2) mean 43 GHz flux density and standard deviation in $\langle S \rangle$ from \cite{Weaver_2022}, (3) mean 43 GHz flux density and standard deviation from this study, (4) mean distance and standard deviation in $\langle r \rangle$ from \cite{Weaver_2022}, and (5) mean distance and standard deviation from this study.}
\end{table}

\section{Synchrotron cooling versus adiabatic cooling} \label{App1}
\begin{figure}[h]
    \centering
    \includegraphics[width=\columnwidth]{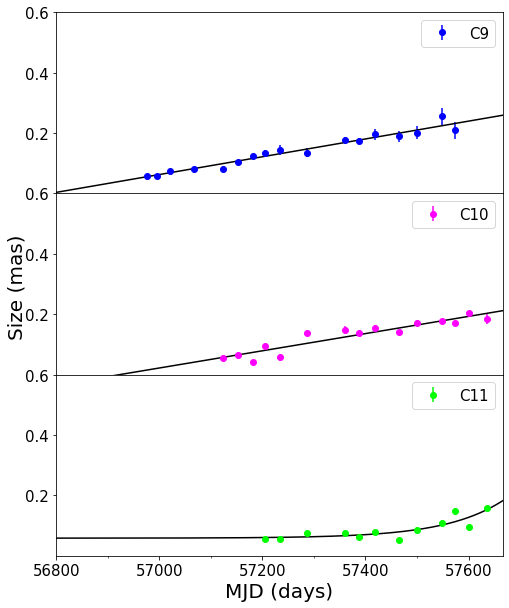}
    \caption{Each panel presents the results of the power-law fitting of the size evolution of each component over time. The panels, from top to bottom, show the results for C9, C10, and C11. The sizes, determined using FWHM values, are shown as scatter points with error bars, while the best power-law fits are represented by solid black lines.} \label{fig:fig8}
    \vspace{5mm}
\end{figure}

When investigating the magnetic field strength based on the cooling timescale, we assumed that the flux density decay is attributed to synchrotron cooling. However, in a relativistic jet, the particle energy density may decrease due to jet expansion, leading to adiabatic cooling.
To ascertain whether this decrease was caused by adiabatic expansion or radiation, we investigated the time variation at size $d$ (or the volume) of the jet components by fitting an exponential function to the data, yielding a jet expanding timescale of $\tau_d$.
The fitting results for each component are shown in Fig.~\ref{fig:fig8}.
The expansion timescale of the source size was found to be $\tau_d\sim10^6$~days, which is significantly longer than the cooling timescales of $\tau_\text{cool}=43$–$222$~days, as discussed in Section \ref{3.4}.
This implies that the flux density decay is primarily attributed to synchrotron cooling rather than adiabatic cooling.

\end{appendix}

\bibliographystyle{aa}          
\end{document}